\documentclass[11pt,superscriptaddress]{article}

\usepackage[a4paper, total={5.8in, 9in}]{geometry}
\usepackage[indent,skip=4pt]{parskip}
\usepackage{amsmath,amssymb,amsfonts,amsthm,mathrsfs,mathtools,cases}
\usepackage{dsfont}
\usepackage{authblk}
\usepackage[colorlinks=true,linkcolor=blue, citecolor=blue, bookmarks]{hyperref}
\usepackage{cleveref}
\usepackage{microtype}
\usepackage{comment}
\usepackage{graphicx,adjustbox} 
\usepackage{subfig}
\usepackage{wrapfig}
\usepackage[font=small,labelfont=bf]{caption}
\usepackage[usenames,dvipsnames]{xcolor}  
\usepackage{booktabs}
\usepackage{braket}
\usepackage[nottoc]{tocbibind}
\usepackage{nicematrix}
\usepackage{tikz}
\usepackage{pgfplots}
\usepackage{cite}
\pgfplotsset{compat=1.18}
\usepackage{enumitem}

\usetikzlibrary{matrix}
\usetikzlibrary{decorations.markings}
\usetikzlibrary{patterns}
\usetikzlibrary{arrows.meta}

\numberwithin{equation}{section}

\crefname{figure}{Fig.}{Figs.}
\crefname{equation}{Eq.}{Eqs.}
\crefname{section}{Sec.}{Secs.}
\crefname{appendix}{Appendix}{Appendices}
  
\colorlet{darkerblue}{MidnightBlue!20!black}
\colorlet{lightblue}{blue!70!white}

\hypersetup{
    colorlinks,
    linkcolor={Blue},
    citecolor={Blue},
    urlcolor={Blue},
    backref=true
}

\newcommand{\beq}{\begin{equation}}
\newcommand{\eeq}{\end{equation}}
\newcommand{\beqa}{\begin{eqnarray}}
\newcommand{\eeqa}{\end{eqnarray}}
\newcommand{\vk}{ \boldsymbol{k}}
\newcommand{\vx}{ \boldsymbol{x}}
\newcommand{\R}{{\mathbb{R}}}

\usepackage[framemethod=TikZ]{mdframed}

\title{\bf Quasiparticle Picture for Entanglement Hamiltonians  in Higher Dimensions}

\author{Riccardo Travaglino$^1$, Colin Rylands$^1$ and Pasquale Calabrese$^{1,2}$}

\date{}

\begin{document}

\maketitle
{\small
\vspace{-5mm}  \ \\
{$^{1}$}  SISSA and INFN Sezione di Trieste, via Bonomea 265, 34136 Trieste, Italy\\[-0.1cm]
\medskip
{$^{2}$}  International Centre for Theoretical Physics (ICTP), Strada Costiera 11, 34151 Trieste, Italy\\[-0.1cm]
\medskip
}

\begin{abstract}
We employ the quasiparticle picture of entanglement evolution to obtain an effective  description for the out-of-equilibrium Entanglement Hamiltonian at the hydrodynamical scale following quantum quenches in free fermionic systems in two or more spatial dimensions. 
Specifically, we begin by applying dimensional reduction techniques in cases where the geometry permits, building directly on established results from one-dimensional systems.
Subsequently, we generalize the analysis to encompass a wider range of geometries. 
We obtain analytical expressions for the entanglement Hamiltonian valid at the ballistic scale, which reproduce  the known quasiparticle picture predictions for the Renyi entropies and full counting statistics. 
We also numerically validate the results with excellent precision  by considering quantum quenches from several initial configurations.

\end{abstract}
\newpage

\tableofcontents

\section{Introduction}
The study of out-of-equilibrium quantum many body systems has been one of the most interesting and fecund areas of research in quantum statistical mechanics in the last few decades. Driven by an unprecedented interplay between novel experimental techniques~\cite{Kinoshita2006AQN, outofeq_exp1,outofeq_exp2,outofeq_exp3,outofeq_exp4} and  deep theoretical insights~\cite{PhysRevA.43.2046,srednicki1,Deutsch_2018,Rigol:2007juv,Rigol}, a more profound understanding of non-equilibrium quantum systems has emerged~\cite{RevModPhys.83.863,DAlessio:2015qtq,Gogolin_2016}. A central theme of this narrative revolves around the local relaxation of many quantum systems undergoing unitary time evolution and the main protagonist of this is the reduced density matrix of a particular subsystem, 
\begin{eqnarray} \rho_A(t)=\operatorname{Tr}_{\bar A}[\rho(t)].
\end{eqnarray}
Here, $A$ is the subsystem of interest, $\bar{A}$ is its  complement and $\rho(t)$ is the density matrix of the full system at time $t$ which we take to have been quenched far from equilibrium. That is, the state is given by 
\begin{eqnarray}
    \rho(t)=e^{-iHt}\ket{\psi}\bra{\psi}e^{i H t}
\end{eqnarray} 
with $\ket{\psi}$ being the initial state of the system and $H$, the Hamiltonian. After substantial efforts using analytic and numerical techniques, the prevailing view is that at long time, barring exceptional circumstances~\cite{abanin2019many},  $\rho_A(t)$ approaches a stationary state, whose properties can be described using statistical mechanics. This stationary state is given by a Gibbs ensemble which incorporates all conserved charges of the dynamics, e.g., in a chaotic system only $H$ will be conserved whereas in an integrable system an extensive number of conserved charges should be taken into account~\cite{calabrese2016introduction,vanicat2018integrable, VidmarRigol, essler2016quench, doyon2020lecture, bastianello2022introduction, alba2021generalized}.  Typically, this relaxation to the steady state is probed indirectly, through the computation of local observables or more complicated quantities like the entanglement entropy. A complete understanding, however,  can only be gained by directly studying $\rho_A(t)$ itself or, more conveniently, the entanglement Hamiltonian $K_A$. This is defined to be
\begin{equation}
      \rho_A = \frac{1}{\mathcal
 Z_A}e^{ -K_A}, \hspace{0.3cm} \mathcal{Z}_A = \operatorname{Tr}e^{ -K_A},
\end{equation}
where the definition is meaningful since the reduced density matrix is hermitian and positive semi-definite. The highly complicated nature of $\rho_A(t)$ makes this a formidable task, however, in certain scenarios remarkable simplifications can occur. For example, quantum field theories can rely upon the powerful 
Bisognano-Wichmann theorem which states that the entanglement Hamiltonian for ground states of critical systems has simple, local, few body form \cite{Bisognano:1975ih,Bisognano:1976za}. These results can sometimes be extended to nonequilibrium systems~\cite{Cardy_2016} but few other results for the quench dynamics of $K_A(t)$ have been obtained~\cite{digiulio2019entanglement,zhu2020entanglement} and a systematic approach to its calculation has been lacking.    

For integrable models evolving after a quantum quench, the most widely applicable theory for studying the local relaxation to the stationary state  is the quasiparticle picture (QPP)~\cite{quench2, alba1,alba2}. This effective theory, in its simplest form, posits that the quench produces correlated pairs of long lived quasiparticles with opposite momenta. These pairs are produced locally at each point in space and correlations exist only between quasiparticles produced at the same point and with opposite momenta. As these pairs propagate throughout the system, they spread their initial correlations to separate spatial regions and armed with some data concerning the initial correlations one can make quantitative predictions about the system dynamics.
The picture can therefore be viewed as a semiclassical effective theory which emerges in the scaling limit of large subsystem sizes and long times. The intuitive and simple nature of the description belies its rigor. Indeed, the method can exactly describe the entanglement entropy growth after quantum quenches in $(1+1)$-dimensional CFT where it was originally developed~\cite{quench2} and  also be shown to emerge from exact calculations in non-interacting spin chains~\cite{fagotti2008evolution}.  The applicability of the QPP extends beyond entanglement entropy in free models, and can be used not only to study entanglement in interacting integrable systems in one spatial dimension~\cite{alba1,alba2,alba2017quench,alba2019entanglement,murciano2022post,PRX} but other quantities also.  These include: correlation functions~\cite{calabrese2012quantum}, entanglement negativity~\cite{coser2014entanglement,alba2019quantum}, full counting statistics~\cite{groha2018full,horvath2024full,bertini2023nonequilibrium}, symmetry resolved entanglement~\cite{bertini2023nonequilibrium,parez2021exact,parez2021quasiparticle}, operator entanglement~\cite{dubail2017entanglement,rath2023entanglement}, and the entanglement asymmetry~\cite{ares2023entanglement,murciano2024entanglement,bertini2024dynamics,rylands2024microscopic}. It can also be modified to account for the presence  of dissipation~\cite{alba2021spreading} or non-local correlations~\cite{bc-18,lagnese2022entanglement,chalas2024quench}.  Thus, an abundance of quantities which probe the nature of the local, many body quantum state $\rho_A(t)$ are amenable to a quasiparticle description. Up until recently, however, it was thought that the complicated nature of the state itself, and more specifically the entanglement Hamiltonian, did not admit such a description. In~\cite{1D} it was shown that, in fact, the entanglement Hamiltonian for free fermion models following a quench from certain initial states in $(1+1)$ dimensions can be derived using the quasiparticle picture. Therein, by using the QPP as an effective theory from the outset, a prediction for $K_A(t)$ in the scaling limit was obtained and subsequently verified using exact numerical simulations. The QPP, therefore, gives a proper understanding of the actual dynamics of these systems rather than being merely a convenient interpretation of results obtained for a specific set of quantities. 

\begin{figure}
    \centering
    \begin{tikzpicture}[scale=0.75]
   
        \draw[black,thick,->] (-1,0) -- (13,0) node[right,scale=1.3]{$x$};
         \draw[black,thick,->] (-1,0) -- (-1,5) node [left, scale=1.3]{$t$};
        \shade [top color=Fuchsia, bottom color=cyan, opacity=0.5]
(8,0) rectangle (13,5);
\shade [top color=Fuchsia, bottom color=cyan, opacity=0.5]
(-1,0) rectangle (4,5);
\shade [top color=yellow, bottom color=red, opacity=0.8]
(4,0) rectangle (8,5);
\draw[thick,black] (4,0)-- (4,5);
\draw[thick,black] (8,0)-- (8,5);
;
\draw[black,dashed,line width=2pt] (7,0)--(8,1) ;
\draw[black,dashed,line width=2pt] (7,0)--(6,1);

\draw[black,line width=2pt,->] (8,1)--(10,3) node[right]{$+k$} ;
\draw[black,line width=2pt,->] (6,1)--(4,3) node[left]{$-k$};

\draw[black,dashed,line width=2pt] (3,0)--(4,0.5) ;
\draw[black,dashed,line width=2pt] (3,0)--(2,.5);
\draw[black,line width=2pt,->] (4,0.5)--(6,1.5)  ;
\draw[black,line width=2pt,->] (2,.5)--(0,1.5);

\node [scale=1.5] at (1.5,5.5){$\overline{A}$};
\node [scale=1.5] at (10.5,5.5){$\overline{A}$};
\node [scale=1.5] at (6,5.5){$A$};
\draw[black,->,line width=2pt,dashed] (6,0)--(7,4) ;
\draw[black,->,line width=2pt,dashed] (6,0)--(5,4);

    \end{tikzpicture}
    \caption{Quasiparticle picture of quench dynamics. After a quench 
 pairs of correlated quasiparticles are produced which propagate through the system.  At each instant of time, dashed lines correspond to pairs which are not shared between $A$ and $\overline{A}$ and therefore do not contribute to the entropy. Solid lines in contrast correspond to shared pairs, which contribute to the entanglement. In the notation of equation \eqref{eq:decomposition}, dashed lines contribute to the pure part of the reduced density matrix, while solid lines contribute to the entangling part.}
    \label{fig:qpp}
\end{figure}
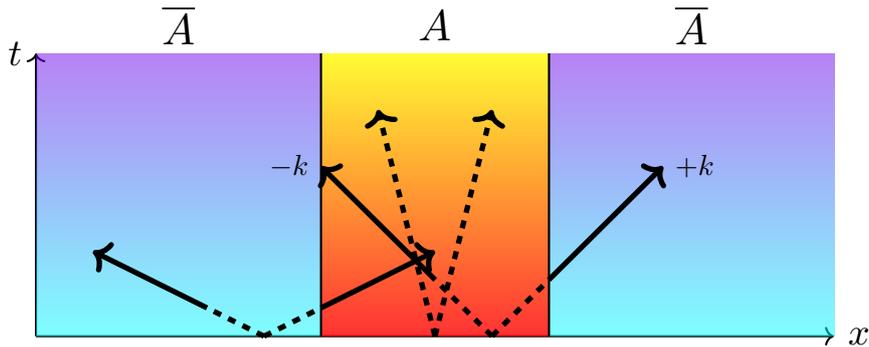

A great deal of attention has been focused on the quench dynamics of $(1+1)$-dimensional systems. Recent efforts, however, have sought to extend our understanding of the entanglement spreading~\cite{Murciano_2020,molly,DimensionalReduction} and symmetry restoration~\cite{shion2} to higher dimensions using a mix of exact methods and the QPP. In this paper we carry on this line of investigation and extend the quasiparticle approach to the entanglement Hamiltonian to $(d+1)$-dimensional systems. In particular, we focus on the quench dynamics of free fermionic lattice systems quenched from two general classes of Gaussian initial states. Using the reasoning of the quasiparticle picture we derive, asymptotically exact and analytic predictions for $K_A(t)$ in a range of quenches. For clarity, we focus much of our attention on two spatial dimensions and confirm our predictions against exact numerical calculations. 

The structure of the paper is as follows. In section \ref{sec:prel} we set up the problem by presenting the Hamiltonians and the type of quenches which will be considered in the following. In section \ref{sec:1d} we then review and expand upon the one-dimensional case, introducing the main features of the hydrodynamic approximation that will be used in the following. Sections \ref{sec:EhDimred} and \ref{sec:gengem} contain the main results of the paper, namely the extensions of quasiparticle approach to higher dimensions, first for strip geometries which can be approached with dimensional reduction techniques and then to generic geometries.  
Several numerical tests for different initial states are presented in Sec. \ref{sec:examples} in support of the analytical claims, before drawing our conclusions in section \ref{sec:concl}. Since in the main text we will restrict for readability to the two-dimensional case and to some specific initial states, in the appendix we will then generalize to arbitrary dimensions and to arbitrary initial states, together with giving additional details on the calculations and proofs of some statements of the main text.

\section{Hamiltonian and quench protocol}
\label{sec:prel}
We consider the quench dynamics of a free fermionic lattice system in $d$ spatial dimensions. The Hamiltonian we shall use is given by
\begin{equation}\label{eq:Hamil}
    H = - \frac{1}{2}\sum_{\braket{\boldsymbol{x},\boldsymbol{x}'}} c^\dagger_{\vx}c_{\vx'},
\end{equation}
where the sum is over nearest neighbors on a hyper-cubic lattice of length $L_i$ in the $i$-th direction.  We take periodic boundary conditions in all directions so that the system geometry is a $d$-dimensional hyper-torus, $\mathcal{S}^d$.   
In momentum space, the Hamiltonian reads 
\begin{equation}
    H = -\sum_{\vk}  \varepsilon_{\vk} c^\dagger_{\vk} c_{\vk},\hspace{0.3cm} \varepsilon_{\vk} = - \sum_{i=1}^d \cos(k_i).  \label{eq:free}
\end{equation}
All of our analysis can, however, be applied more generally to arbitrary dispersion relations $\epsilon(\boldsymbol{k})$ and even continuous models with the salient piece of information being the quasiparticle velocity
\begin{eqnarray}
    \boldsymbol{v}(\vk)=\nabla_{\vk} \epsilon(\vk).
\end{eqnarray} For numerical checks on our results, though, we will use the specific Hamiltonian above, ~\eqref{eq:Hamil}, for which $\boldsymbol{v}(\vk)=(\sin(k_x),\sin(k_y),\dots)$.  Moreover, for some cases, explained further below, we shall make use the toroidal geometry of the system but this can also be relaxed and open boundary conditions in some or all directions  can be treated.

We study the unitary time evolution governed by $H$ from two classes of simple initial states: symmetry breaking and symmetry preserving.  In particular, the first class of states that we will focus on are the so called ``squeezed states", which break the $U(1)$ particle number symmetry. They are given by
\beq
    \ket{\psi} =\mathcal{N} \exp{\Big(\sum_{|\vk|>0}\mathcal{M}(\vk)c_{\vk}^\dagger c_{-\vk}^\dagger\Big)} \ket{0} = \mathcal{N}\exp{\Big( \sum_{\vx,\boldsymbol{y}}\tilde{\mathcal{M}}(\vx-\boldsymbol{y})c^\dagger_{\vx}c^\dagger_{\boldsymbol{y}}\Big)} \ket{0},
    \label{eq:squeezed}
\eeq
where $\mathcal{N}$ is the normalization of the state and $\mathcal{M}(k)$ is an arbitrary odd function of $\vk$, whose Fourier transform is denoted by $\tilde{\mathcal{M}}(\vx)$. The particular form of $\mathcal{M}(\vk)$ is not important for our general treatment but we shall specialize to some specific cases later on. These types of states are ubiquitous in quantum quenches, they appear in all situations for which the pre- and post-quench Hamiltonians are related by Bogoliubov transformations and are sometimes referred to as mass quenches \cite{Fioretto_2010}.
Upon expanding the exponential, we can alternately express the initial state  in the following way
\begin{equation}\label{eq:occupation_function}
    \ket{\psi}= \prod_{\vk}\left( \sqrt{1-n(\vk)} + e^{i\varphi_{\vk} }\sqrt{n(\vk)} c^\dagger_{\vk} c^\dagger_{-\vk}\right)\ket{0},
\end{equation}
 Where we have explicitly incorporated the normalization and introduced the mode occupation function,
\begin{equation}\label{eq:occupation function}
n(\vk)=\bra{\psi}c^\dag_{\vk} c_{\vk} \ket{\psi}= \frac{|\mathcal{M}(\vk)|^2}{1+|\mathcal{M}(\vk)|^2}. 
\end{equation}
as well as a phase $e^{i\varphi_{\vk}}$ which will turn out to be unimportant. Along with the quasiparticle velocity, $n(\vk)$ shall completely determine the form of the entanglement Hamiltonian. 

The second class of states which will be considered preserve the particle number symmetry. They are closely related to the ones investigated in \cite{Bertini_2018,molly}, namely states with shift symmetry $\mu_i$ along the $i$-th axis, which therefore have elementary cells of volume $|\mu| = \mu_1\mu_2...\mu_d$. A quite general state of this form in which there are $\mathtt{n}$ fermions per unit cell is    given by
\begin{equation}\label{eq:symm_preserv_general}
\ket{\psi}= \prod_{\boldsymbol{j}=1}^{\boldsymbol{L}/\boldsymbol{\mu}} \prod_{\lambda=1}^{\mathtt{n}}(\sum_{\boldsymbol{m}_\lambda=0}^{\boldsymbol{\mu}-1}a^{(\lambda)}_{\boldsymbol{m}_\lambda} c_{\boldsymbol{\mu} \boldsymbol{j}-\boldsymbol{m}_\lambda}^\dagger)\ket{0} 
\end{equation}
where we have used the compact notation
 \begin{equation}
\prod_{\boldsymbol{j}=1}^{\boldsymbol{L}/\boldsymbol{\mu}} = \prod_{j_x=1}^{L_x/\mu_x}\prod_{j_y=1}^{L_y/\mu_y}\dots\prod_{j_d=1}^{L_d/\mu_d},\hspace{0.5cm} \sum_{\boldsymbol{m}=0}^{\boldsymbol{\mu}-1} = \sum_{m_x=0}^{\mu_x-1}\sum_{m_y=0}^{\mu_y-1} \dots
 \end{equation}
and we also have expressed $\boldsymbol{\mu}\boldsymbol{j}=(\mu_xj_x, \mu_yj_y,...)$.   Although the states \eqref{eq:symm_preserv_general} might appear rather complicated, they are simply the most general way of writing any shift-symmetric state with a fundamental cell defined by $\boldsymbol{\mu}=(\mu_x,\mu_y,\dots)$. In practice, we will mostly be concerned with states with one or two site shift symmetry, in which there are only one or two fermions per units cell and  for one and two spatial dimensions. A typical example of such a state is the one-dimensional dimer,
 \begin{eqnarray} \label{eq:dimer1d}
\ket{\psi}=\prod_{j=1}^{L/2}\frac{c^\dag_{2j-1}-c^\dag_{2j}}{\sqrt{2}}\ket{0}
\end{eqnarray}
which has two-site shift symmetry, $\mu_x=2$, one fermion per unit cell, $\mathtt{n}=1$ and $a^{(1)}_{0}=a^{(1)}_1=1/\sqrt{2}$.

We aim to compute the post quench entanglement Hamiltonian, $K_A(t)$ emerging from the states above and, ultimately, obtain transparent analytic predictions for its behavior using the QPP. 
Since the Hamiltonian  we use is quadratic in the fermion operators and the states introduced above are Gaussian, we can make use of the powerful and exact analytic tools which are available in such a setting. 
The key ingredient for these is the two-point, fermion correlation function,
 \begin{equation}
     C_{\vx, \vx'}{(t)} = \braket{\psi(t) |c^\dagger_{\vx} c_{\vx'}|\psi(t)},
 \end{equation}
where $\ket{\psi(t)}=e^{-iHt}\ket{\psi}$ is the time evolved state, which is also Gaussian. 
In this case, Wick's theorem can be applied and therefore all information about the correlations in the subsystem $A$ are  encoded in the restricted correlation matrix $C_A(t)$, which is the $|A|\times|A|$ matrix whose elements are given by  $C_{\vx, \vx'}{(t)}$ with $\vx,\vx'\in A$. As a result, one can relate $C_A(t)$ to the entanglement Hamiltonian using the 
 Peschel formula \cite{PhysRevB.69.075111,Peschel_2009}
 \begin{equation}
    C_A(t) = \frac{1}{1+e^{k_A{(t)}}},\label{eq:peschelformula}
\end{equation}
where  $k_A{(t)}$ is the matrix of coefficients of the entanglement Hamiltonian expressed in matrix form as
\begin{equation}\label{eq:peschel}
K_A(t)= \sum_{\boldsymbol{i},\boldsymbol{j}}k_{A,\boldsymbol{i},\boldsymbol{j}}{(t)}c_{\boldsymbol{i}}^\dagger c_{\boldsymbol{j}}.
\end{equation}
This  is an extremely useful result since,  in many instances, the correlation matrix can be evaluated exactly~\cite{Eisler_2007}, allowing one to obtain the exact results for the entanglement Hamiltonian.
However, as often happens, this exact solution is in general extremely complicated and potentially not analytically tractable, since inverting explicitly \eqref{eq:peschelformula} for arbitrary states is a formidable task.  Thus, generically~\eqref{eq:peschel} does not provide great physical insight but instead can be  used as the basis for numerical calculations. In contrast, while the quasiparticle picture approach is only approximate it nevertheless provides an intuitive understanding of the dynamics of $K_A(t)$. We shall employ~\eqref{eq:peschelformula} as a numerical check on the validity of the quasiparticle approach to the entanglement Hamiltonian.

\begin{figure}[h]
\centering
\begin{tikzpicture}
\begin{axis}
[view={20}{55},axis line style={draw=none},  tick style={draw=none}, label style = {draw=none}, yticklabel={\empty}, xticklabel={\empty}, zticklabel={\empty}]
\addplot3
[domain=0:360,y domain=-40:260,
variable=\u,variable y=\v,
samples=40,z buffer=sort,
surf,colormap/cool,shader=interp,opacity=0.8]
({(3+cos(u))*cos(v)},
{(3+cos(u))*sin(v)},
{sin(u)});

\addplot3
[domain=0:360,y domain=260:320,
variable=\u,variable y=\v,
samples=40,z buffer=sort,
surf,colormap/redyellow,shader=interp,opacity=0.8]
({(3+cos(u))*cos(v)},
{(3+cos(u))*sin(v)},
{sin(u)});
\end{axis}

\draw [black, fill] (3.4,2) node [scale=1.3] {$A$};
\draw [black, fill] (3.4,4) node [scale=1.3] {$\overline{A}$};
\draw [<->] (2.2,0.8) to[out=-10,in=190] (4.7,0.8);
\draw [black] (3.4,0.5) node [scale=0.7] {$\ell$};

\end{tikzpicture}
\caption{Two dimensional realization of the strip geometry which allows for the use of dimensional reduction.  }
\label{fig:dr}
\end{figure}
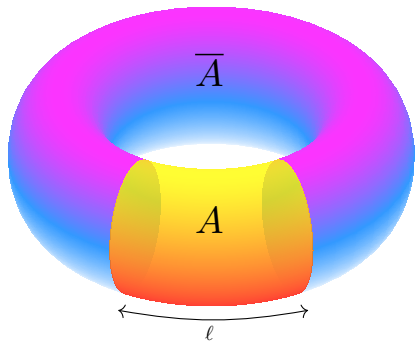
As intimated above, even for free theories, studying systems in dimensions $d>1$ via analytic means can be rather challenging.
A powerful tool to study the problem in a simplified fashion is provided by dimensional reduction. Therein, one employs a particularly convenient geometry of both the system and subsystem which conspire, along with the symmetries of the model, to effectively reduce the number of spatial dimensions. This approach has proven quite successful previously in illuminating aspects of the entanglement entropy in higher dimensions~\cite{PhysRevB.62.4191, Murciano_2020, DimensionalReduction,shion2,yamashika2024quenching}.  With this in mind, we can take our subsystem of interest, $A$, to be  a strip which covers entirely all dimensions but one, 
\begin{equation}
 A = [1,\ell] \times \mathcal{S}^{d-1},
 \label{eq:geometry}
\end{equation}
where $\ell<L$, as shown in Figure \ref{fig:dr} for 2-dimensions.
For geometries of this type, the time evolution of entanglement entropy was studied in detail in  \cite{DimensionalReduction, shion2}, and it was shown that the results are consistent with a convenient adaptation of the quasiparticle picture.
In fact, the specific geometry allows for a great simplification of the problem of evaluating the bipartite entropy of $A$ since it allows to decouple the system in a direct sum of 1-dimensional systems.  
For example, a two dimensional system in which the initial state is chosen of the form \eqref{eq:symm_preserv_general} with $\mu_y=1$, namely 1-site translational invariance along y. The correlation matrix, up to a unitary transformation, can then be expressed as a direct sum
\begin{equation}\label{eq:dime_red_corr}
    C_{\vx,\vx'}  \sim \bigoplus_{n_y=0}^{L_y-1}(C_{q_y})_{x,x'},
\end{equation}
where the correlation matrices $C_{q_y}$ are constructed in terms of the mixed real-momentum space operators, obtained by performing a Fourier transform only on the y-direction,
\begin{equation}
    c_{x,q_y} = \frac{1}{\sqrt{L_y}} \sum_y  e^{iyq_y}c_{\vx},
    \label{eq:partialfourier}
\end{equation}
and the momenta along the y direction is quantized as $q_y = \frac{2 \pi}{L_y} n_y$. The algebraic structure exhibited by~\eqref{eq:dime_red_corr} then has consequences for the entanglement entropy which can be written as a sum over different $q_y$ contributions. As we show below this  is also for $K_A(t)$ itself which admits a similar decomposition in terms of $q_y$ modes.

\section{One dimensional case }
\label{sec:1d}We begin by providing an  overview of the quasiparticle picture approach to determining the entanglement Hamiltonian in one spatial dimension. Here we closely follow and expand upon the presentation of~\cite{1D} and~\cite{bertini2018entanglement} which directly employ the QPP from the outset rather than seeing it emerge from microscopic calculations. To this end, we should recall that the QPP is an effective theory which captures the quench dynamics in the thermodynamic limit and for large subsystem sizes and long times. In particular, we should take $L\gg \ell$ and $L,\ell, t\to \infty$ with $t/\ell\equiv\zeta $ held fixed. This ballistic scaling limit, is the same as the one in which Euler scale hydrodynamics is expected to occur in integrable models~\cite{GHD1,GHD2} and is the one which we shall employ throughout this section. 

\subsection{Setup}
We consider a lattice system characterized by the one dimensional tight binding  Hamiltonian,
\begin{equation}
    H = -\frac{1}{2} \sum _{i=1}^L c_i^\dagger c_{i+1} + h.c. = -\sum_{k}\cos ( k) c_k^\dagger c_k.
\end{equation}
with the quasiparticle velocity being $v(k)=\sin(k)$. To start with, we consider the initial state to be the squeezed state given by
\beq
    \ket{\psi} =  \frac{1}{\mathcal{N}}\exp{\Big (\sum_{x,x'=1}^L \tilde{\mathcal{M}}(x-x')c^\dagger_{x}c^\dagger_{x'}\Big)}\ket{0},
\eeq
with $\tilde{\mathcal{M}}(x)$ an odd function of $x$. 
In addition, since there is only a single spatial dimension, we take the subsystem to be simply a single block of $\ell$ contiguous sites, $A=[1,\ell]$ starting, without loss of generality, at the first site. Since we will use the quasiparticle picture which is effective in the ballistic scaling limit, we adopt a  hydrodynamic approach, by dividing the full system into fluid cells of size $\Delta$, such that $1 \ll \Delta \ll \ell$.  Furthermore, we split the lattice coordinate into two parts, $x=x_0+\varkappa$,  where $x_0\in [1,L/\Delta]$ indexes the fluid cell to which  $x$ belongs and $\varkappa\in[0,\Delta-1]$ gives the position inside the fluid cell. We now impose some conditions on the form of our initial state and restrict to cases where the correlation length of the state, denoted by $\xi$, is much shorter than the fluid cell size. Practically this means we impose that 
\begin{eqnarray}
    \label{eq:fluid_cell}\tilde{\mathcal{M}}(x)\approx 0,~~\forall |x|>\xi,~\xi\ll\Delta .
\end{eqnarray}
This type of restriction does not overly limit our choices of initial state. Indeed, as stated in the previous section, ground states of massive free models generically take the form of a squeezed state, in which case they should naturally have a finite correlation length, and obey~\eqref{eq:fluid_cell}. 
As a consequence of this, the initial state can be simplified to 
\begin{eqnarray}
    \ket{\psi}&=&\frac{1}{\mathcal{N}}\exp{\Big (\sum_{x_0=1}^{L/\Delta}\sum_{\varkappa,\varkappa'=0}^{\Delta-1} \tilde{\mathcal{M}}(\varkappa-\varkappa')c^\dagger_{x_0+\varkappa}c^\dagger_{x_0+\varkappa'}\Big)}\ket{0}\\\label{eq:fluid_initial_state}
    &=&\frac{1}{\mathcal{N}}\prod_{x_0=1}^{L/\Delta} \exp{\Big (\sum_k\mathcal{M}(k)b^\dagger_{x_0,k}b^\dagger_{x_0,-k}\Big)}\ket{0}
\end{eqnarray}
where in the second line we have introduced the creation operators, 
\begin{equation}
    b^\dagger_{x_0,k}=\frac{1}{\sqrt{\Delta}} \sum_{\varkappa=0}^{\Delta-1} e^{-ik\varkappa} c^\dagger_{x_0+\varkappa} 
    \label{eq:fluidcelloperators},
\end{equation}
which have been partially Fourier transformed inside the fluid cell, and also assumed a tensor product structure over the different  cells.  Here, the fluid cell momenta run over a set of values quantized according the fluid cell size $\Delta$, i.e. $k\in \frac{2\pi}{\Delta}\mathbb{Z}_\Delta$ where $\mathbb{Z}_\Delta$ denotes the integers modulo $\Delta$. We note, however, that in doing so we only select a basis for the space inside the cell and do not impose un-physical periodic boundary conditions on the cell itself~\cite{granet2023wavelet}.

The expression, \eqref{eq:fluid_initial_state} allows us to write the density matrix in a simplified fashion as
\begin{equation}
    \rho{(0)} = \prod_{x_0}\prod_{k>0} \rho_{x_0,k}
\end{equation}
where $\rho_{x_0,k}$ is the density matrix of a pair of quasiparticles of opposite momenta inside  the cell located at $x_0$. Given the pair structure, we need only take the product over $k>0$ to avoid over counting. To express this in a convenient form, we represent the vacuum density matrix as 
\begin{equation}
\ket{0}\bra{0}=\prod_{x_0}\prod_{k}(1-\hat{n}_{x_0}(k))
\end{equation}
where we introduced $\hat{n}_{x_0}(k)=b^\dag_{x_0,k}b_{x_0,k}$ from which we find that 
\beqa\nonumber
    \rho_{x_0,k} &=& 
    n(k) \hat{n}_{x_0}(k)\hat{n}_{x_0}(-k)   + (1-n(k))(1- \hat{n}_{x_0}(k))(1- \hat{n}_{x_0}(-k)) \\ &+&  \sqrt{n(k)(1-n(k))}(e^{i\varphi_k}b_{x_0,k}^\dagger b_{x_0,-k}^\dagger +e^{-i\varphi_k}  b_{x_0,-k}b_{x_0,k}).
    \label{eq:densitymatrix}
\eeqa
Here, the occupation functions $n(k)$ are defined as in~\eqref{eq:occupation_function} and $e^{i\varphi_k}=\mathcal{M}(k)/|\mathcal{M}(k)|$. 
Note that, the occupation functions $n(k)$ do not acquire a spatial dependence because the quench is homogeneous. For  quenches from inhomogeneous states which vary slowly on scales larger than $\Delta$ we could take $n(k)\to n(k,x_0)$ and a similar analysis could be carried out. We leave these finer details for future work.

\subsection{Quasiparticle dynamics}
Having expressed the initial density matrix in a manner which is amenable to a hydrodynamic treatment, we now proceed to study the time evolution using the QPP. Within this framework, we treat the $b^\dag_{x_0,k}$ as being creation operators for a set of semiclassical quasiparticle modes which propagate ballistically throughout the system.  We approximate their actual evolution by~\cite{1D},
\begin{equation}\label{eq:QPP_heisenber_evolution}
e^{iHt}b^\dag_{x_0, k} e^{-iHt} \approx b^\dag_{x_0+ v(k)t, k}.
    \end{equation}
 Thus, the quasiparticles are  coherently transported between fluid cells according to their velocity $v(k)$ from the cell labeled by $x_0$ to the one labeled by $x_t(k)=x_0+v(k) t$ and any dispersive effects are assumed to be subleading on length scales comparable to $\Delta$. We should stress that the choice of $b^\dag_{x_0,k}$ is not devoid of subtleties. For example, we could make Bogoliubov transformation on the fluid cell operators and treat them as the propagating quasiparticles instead. Typically, such a rotation will lead to incorrect results, however there are exceptional cases where this is necessary due to the  presence of an extra set of nontrivial  conserved quantities~\cite{ares2023lack}. In this section we avoid such situations but note they can be treated in the same way through the proper choice of quasiparticle.

 The time evolved density matrix for a single pair originating from $x_0$ thus reads,
 \beqa\nonumber
    \rho_{x_0,k}(t) \!&=&\! 
    n(k) \hat{n}_{x_t}(k)\hat{n}_{x_t}(-k)    + (1-n(k))(1- \hat{n}_{x_t}(k))(1-\hat{n}_{x_t}(-k)) \\ &&+  \sqrt{n(k)(1-n(k))}(e^{i\varphi_k}b_{x_t(k),k}^\dagger b_{x_t(-k),-k}^\dagger +e^{-i\varphi_k} b_{x_t(-k),-k}b_{x_t(k),k}).
    \label{eq:densitymatrix}
\eeqa
where $ \hat{n}_{x_t}(\pm k)=b^\dag_{x_t(\pm k),\pm k}b_{x_t(\pm k),\pm k}$. The members of each pair originating at $x_0$ are therefore transported to different fluid cells over time. 

We are interested in calculating the reduced density matrix inside the subsystem $A=[1,\ell]$. With this in mind, suppose that the right moving particle, $b^\dag_{x_t(k),k}$ is inside the subsystem at time $t$, i.e. $x_t(k)\in A$,  whereas the left moving particle is not, 
 $x_t(-k)\notin A$. In order to obtain the reduced density matrix we must trace out the left moving quasiparticle. Upon taking the trace over the Hilbert space associated to  $b^\dag_{x_t(-k),-k}$ in~\eqref{eq:densitymatrix} and denoting the resulting state by  $\rho_{A,x_0,k}(t) $ we find that, 
\begin{eqnarray}\label{eq:formweneed}
    \rho_{A,x_0,k}(t) &=&n(k) \hat{n}_{x_t}(k) + (1-n(k))(1-\hat{n}_{x_t}(k))\\
    &=& \frac{1}{1+e^{-\eta(k)}}\exp{\big(-\eta(k)\hat{n}_{x_t}(k)\big)}
\end{eqnarray}
where  in the second line we have introduced 
\begin{equation}\label{eq:eta}
    \eta(k) = \log\left(\frac{1-n(k)}{n(k)}\right).
\end{equation}
Note that the phase factor $e^{i\varphi_k}$ has dropped out as it only appears in the off-diagonal terms and only the dependence on $n(k)$ is present.   
Assuming instead that it is the left moving quasiparticle which remains inside the subsystem, we would arrive at a similar expression,
\begin{eqnarray}
    \rho_{A,x_0,-k}(t)  =\frac{1}{1+e^{-\eta(k)}}\exp{\big(-\eta(k)\hat{n}_{x_t}(-k)\big)}.
\end{eqnarray}
We can then carry out this for all modes and fluid cells after which we determine that $\rho_A(t)$ takes the following form,
\begin{equation}\label{eq:decomposition}
    \rho_A(t) \approx \rho_{\rm mixed}(t)\otimes \rho_{\text{pure}}{(t)},
\end{equation}
which is split into a mixed state part containing the entanglement Hamiltonian and a pure part\footnote{Note that this is an effective description based on the quasiparticle picture and not an exact statement regarding the state at all times. In particular, the tensor product in \eqref{eq:decomposition} makes sense only in the hydrodynamic framework, since it is defined over the values of $x_0$ labeling the cells, and arises because each cell is treated independently.}.    To determine each of these separate parts we proceed as follows.   For each fluid cell $x_0$, and momentum $k$, we evolve the corresponding density matrix $\rho_{x_0,k}$ according to~\eqref{eq:densitymatrix} up to a time $t$.  We then check if the new positions of the quasiparticles are inside $A$ or not.   If both quasiparticles are in $A$, neither needs to be traced out and they  contribute to $\rho_{\rm pure}(t)$ as in~\eqref{eq:densitymatrix}. If instead only one member of the pair is inside $A$, then its corresponding partner must  be traced out and the remaining quasiparticle contributes instead to the mixed part as in~\eqref{eq:formweneed}. If neither quasiparticle is inside $A$, then both are traced out and they do not contribute to $\rho_A(t)$. Thus, we have that 
\begin{eqnarray}\label{eq:pure}
    \rho_{\rm pure}(t)&=&\prod_{k>0}\prod_{\{x_0|x_{t}(k)\in A\,\& \, x_{t}(-k)\in A\}}\rho_{x_0,k}(t),\\\label{eq:mixed}
    \rho_{\rm mixed}(t)&=&\prod_{k}\prod_{\{x_0|x_{t}(k)\in A\,\& \, x_{t}(-k)\notin A\}}\rho_{A,x_0,k}(t).
\end{eqnarray}
Note that, in the first line the product is over only positive momenta while in the second there is no such restriction. The mixed part can then be expressed in terms of the  entanglement Hamiltonian 
\begin{eqnarray}
    \rho_{\rm mixed}(t)&=&\frac{1}{
\mathcal{Z}_A}e^{-K_{A,QP}(t)},\\\label{eq:ent_ham_mixed}
    K_{A,QP}(t)&=&\sum_{k}\sum_{\{x_0|x_{t}(k)\in A\,\& \, x_{t}(-k)\notin A\}}\eta(k)\hat{n}_{x_t}(k)
\end{eqnarray}
where $\mathcal{Z}_A=\operatorname{Tr}[e^{-K_{A,QP}}]$ and we use the subscript $QP$ to remind us that this expression is valid within the confines of the QPP. Naturally, the two sets which determine the quasiparticle content of the mixed and pure parts are nonintersecting and change over time. In fact on time scales much longer than the subsystem size, $\zeta=t/\ell\to\infty$, it is not possible for a pair to remain completely inside $A$, meaning that 
\begin{eqnarray}
  \lim_{\zeta\to\infty}  \{x_0|x_{t}(k)\in A\,\& \, x_{t}(-k)\in A\}\to \emptyset
\end{eqnarray}
In contrast, the set of pairs contributing to the mixed part grows in size over time such that 
\begin{eqnarray}
 \lim_{\zeta\to\infty}  \big| \{x_0|x_{t}(k)\in A\,\& \, x_{t}(-k)\notin A,\forall k\}\big|\to\ell.
\end{eqnarray}
Thus, the dimension of the Hilbert space of quasiparticles associated to the mixed state grows so that it eventually saturates to $2^\ell$ while the one corresponding to the pure state shrinks and becomes the null space. These two limiting cases occur concomitantly and once that happens the local state is said to have completely relaxed to its stationary state. 

Typically, the entanglement Hamiltonian is presented in real space only, as opposed to the mixed real space Fourier representation of~\eqref{eq:ent_ham_mixed}. To bring it to the desired form we should perform the inverse Fourier transform within the fluid cells. Moreover, since we are working within the ballistic scaling regime we can simplify our expressions by replacing the sums over lattice sites with integrals, i.e.  $\sum_{\{x_0|x_{t}(k)\in A\,\& \, x_{t}(-k)\notin A\}}\to \int_{0}^\ell {\rm d}x \mathcal{\chi}(x,t)$, where the integration variable is shifted to the coordinate of the right/left mover (depending on which one is in $A$) and $\mathcal{\chi}(x,t)$ is a suitable counting function which selects the shared pairs.
As a result we find that the time evolved QPP entanglement Hamiltonian, re-expressed in terms of the original fermionic operators is~\cite{1D},
\begin{equation}
    K_{A,QP}{(t)} = \int_0^\ell dx \int_{x-z\,\in \,A}\hspace{-20pt} dz \left[\mathcal{K}_R(x,z;t) + \mathcal{K}_L(x,z;t)\right]c_x^\dagger c_{x-z}, 
    \label{eq:1deh}
\end{equation}
where the two kernels distinguish right and left movers and are given by 
\beqa\label{eq:right_kernel}
    \mathcal{K}_R(x,z;t) &=& \int_{k>0}\frac{dk}{2\pi}\eta(k) \Theta(\min(2v(k)t,\ell)-x)e^{-ik z},\\\label{eq:left_kernel}
     \mathcal{K}_L(x,z;t) &=& \int_{k<0}\frac{dk}{2\pi}\eta(k) \Theta(\max(l+2v(k)t,0)+x)e^{-ik z},
\eeqa
and the Heaviside functions which account for the  quasiparticle pairs are shared between $A$ and $\bar A$ and exhibit a light cone structure. In addition, the domain of integration for the $z$ variable ensures that only operators inside $A$ appear in the entanglement Hamiltonian. Equations~\eqref{eq:1deh}, \eqref{eq:right_kernel} and \eqref{eq:left_kernel} are the main result of~\cite{1D}. Along with the expression for $\rho_{\rm pure}(t)$ they give a complete analytic characterization of the reduced density matrix of a quenched free fermion model. They are valid within the limitations of the quasiparticle picture and are expected to hold asymptotically exactly in the  scaling limit. 

Many physical properties of the system can be discerned from examining the form of the entanglement Hamiltonian, $K_{A,QP}(t)$. One example is the linear growth of entanglement from area law to volume law after a quantum quench which is reflected in the locality of the entanglement Hamiltonian. It is known that entanglement Hamiltonians for ground states of gapped systems, whose entanglement obeys an area law, exhibit a local structure \cite{EH}. In the same spirit, the expression \eqref{eq:1deh} also unveils very intuitively the transition between the area law and volume law. For small $t$, the kernels, $\mathcal{K}_{L,R}$ are different from zero only in a region which is contained in light cones originating from the boundary points.  Therefore, $K_A(t)$ is essentially localized at the boundary. As $t$ grows we have a spreading of the light cones which eventually make the entanglement Hamiltonian nonzero in all the system giving rise to  the volume law scaling of entanglement entropy at long times.

\subsection{Symmetry preserving states}
The discussion of the previous section started from the symmetry breaking, squeezed state however we saw that the final result for $K_{A,QP}(t)$ only depended upon the fermion occupation function while  symmetry breaking terms, which could depend upon the phase $e^{i\varphi_k}$ were absent.  As this suggests, the result obtained in ~\cite{1D} is actually much more general. To see this we consider states of the form of \eqref{eq:symm_preserv_general} with two site translational symmetry and one fermion per unit cell. In Fourier space this reads
\begin{equation}\label{eq:symm_preserve_state_1d}
    \ket{\psi}=\prod_k (\sqrt{n(k)} c^\dagger_k +e^{i\varphi_k}\sqrt{1-n(k)} c^\dagger_{-k})\ket{0}.
\end{equation}
A particular example of this is the dimer state~\eqref{eq:dimer1d} wherein $n(k)=\frac{1-\cos(k)}{2}$. We can re-express this in terms of fluid cell modes $b^\dag_{x_0,\pm k}$ as we did previously and then construct the density matrix which factorizes over fluid cells. In this instance, the initial state is a product state and the decomposition into fluid cells is exact, in contrast to the squeezed state. Upon evolving this in time according to the QPP~\eqref{eq:QPP_heisenber_evolution}, we have that
\beqa
\rho(t)&=&\prod_{k>0}\prod_{x_0}\rho_{x_0,k}(t)\\\nonumber
    \rho_{x_0,k}{(t)} &=&n(k)\hat{n}_{x_t}(k)(1-\hat{n}_{x_t}(-k)) + (1-n(k))(1-\hat{n}_{x_t}(k)) \hat{n}_{x_t}(-k) \\\label{eq:symm_presrve_pure} &&+ \sqrt{n(k)(1-n(k)) }\left(e^{i\varphi_k}b_{x_t(k),k}^\dagger b_{x_t(-k),-k}+e^{-i\varphi_k}b_{x_t(-k),-k}^\dagger b_{x_t(k),k}\right).
\eeqa
 This expression is different from \eqref{eq:densitymatrix}, however, they both lead to the same pair structure once we perform the tracing out.  For example, if the left mover is outside of $A$ but the right mover is still inside, we obtain
\beqa
    \rho_{A,x_0,k}{(t)} &=&
   n(k) \hat{n}_{x_t}(k) + (1-n(k))(1-\hat{n}_{x_t}(k)) \label{eq:itworks2}
\eeqa
which is exactly of the same form as \eqref{eq:formweneed}.
Hence, the entanglement Hamiltonian is the same in both cases. It should be stressed, however, that while $K_{A,QP}(t)$ is the same for both initial states, $\rho_A(t)$ are different since the pure part in the symmetry preserving case must be constructed as in~\eqref{eq:pure} but using~\eqref{eq:symm_presrve_pure}.  What we have shown, therefore, is that the two-body structure of the entanglement Hamiltonian predicted in \cite{1D} is a rather general feature of free fermionic one dimensional models, and holds in particular for all initial states with two-site shift symmetry. For other classes of initial states, with different shift symmetry, it is known that the quasiparticle picture can still be applied by considering more complicated structures of correlated multiplets \cite{Bertini_2018}. In these situations, the preceding analysis can be applied as well,  the only complication coming from the counting of the shared multiplets; the main discussion, however, would apply without any conceptual modification.

\subsection{Analytic checks}
The factorized form of the reduced density matrix~\eqref{eq:decomposition}-\eqref{eq:mixed} lends itself to calculating many properties of the system analytically. In this section we show how this can be done for two quantities, the entanglement entropy and full counting statistics.  Expressions for  quantities are known and our derivation of them will serve as an important analytic check of our results. 

We begin by considering the entanglement entropy between $A$ and $\bar A$ defined as,
\begin{equation}
    S_A(t)=-\operatorname{Tr}[\rho_A(t)\log\rho_A(t)].
\end{equation}
To compute this using the results of the previous sections, we note that the pure part will not contribute to the entropy while the product nature of the mixed part allows us to express it as
\begin{eqnarray}
    S_A(t)&=&-\sum_{k}\sum_{\{x_0|x_{t}(k)\in A\,\& \, x_{t}(-k)\notin A\}}\operatorname{Tr}[\rho_{A,x_0,k}(t)\log \rho_{A,x_0,k}(t)],\\
    &=&-\int \frac{dk}{2\pi}{\rm min}(2|v(k)|t,\ell)[n(k)\log n(k)+(1-n(k))\log (1-n(k))]
\end{eqnarray}
where in going to the second line we have used the fact that the contribution to the entanglement entropy given by a density matrix of the for $e^{-\eta(k)\hat{n}(k)}$ is equal to its thermodynamic entropy and also taken the thermodynamic limit. This matches the known expressions for $S_A(t)$
 in free fermion models~\cite{fagotti2008evolution}. 
 
Evidently the entanglement entropy only depends upon the entanglement Hamiltonian, however there are many quantities which also depend upon $\rho_{\rm pure}(t)$. One such quantity is the full counting statistics of the fermion number,
\begin{eqnarray}
    Z_A(\alpha,t)=\operatorname{Tr}[e^{\alpha \hat{N}_A}\rho_A(t)]
\end{eqnarray}
where $\hat{N}_A=\sum_{x\in A}c^\dag_xc_x$ is the charge inside $A$. This again can be calculated straightforwardly using the factorization of~\eqref{eq:decomposition}. In particular, 
\begin{eqnarray}\nonumber
    \log Z_A(\alpha)&=&\sum_{k>0}\sum_{\{x_0|x_{t}(k)\in A\,\& \, x_{t}(-k)\in A\}}\log \operatorname{Tr}[e^{\alpha [\hat{n}_{x_t}(k)+\hat{n}_{x_t}(-k)]}\rho_{x_0,k}(t)]\\&&+\sum_{k}\sum_{\{x_0|x_{t}(k)\in A\,\& \, x_{t}(-k)\notin A\}}\log \operatorname{Tr}[e^{\alpha \hat{n}_{x_t}(k)}\rho_{A,x_0,k}(t)]
\end{eqnarray}
where the first sum is the contribution of the pure part and the second, that of the mixed part. In the thermodynamic limit this becomes 
\begin{eqnarray}\nonumber
     \log Z_A(\alpha)&=&\int _{k>0}\frac{dk}{2\pi}[\ell-{\rm min}(2|v(k)|t,\ell)]\log(1-n(k)+n(k)e^{-2\alpha})\\&&+\int\frac{dk}{2\pi}{\rm min}(2|v(k)|t,\ell)\log(1-n(k)+n(k)e^{-\alpha})
\end{eqnarray}
which matches the known expression~\cite{bertini2024dynamics}.

\section{Two dimensional case}
\label{sec:EhDimred}
Having reviewed how one can obtain the reduced density matrix of a quenched system of free fermions in one dimension through the QPP, we now move on to consider the two dimensional case. In this section we choose a subsystem which is finite in the $x$-direction but entirely covers the $y$-direction, i.e.
\begin{equation}
    A = [1,\ell]\times \mathcal{S},
\end{equation}
 and furthermore we will focus on the class of symmetry preserving states~\eqref{eq:symm_preserv_general}. As in the one dimensional system, we are interested in a particular scaling limit in which $L_x,t,\ell\to\infty$ with $\zeta=t/\ell$ held fixed. We allow, however, for $L_y$ to be finite. This geometry and choice of initial states coincides with those considered in~\cite{DimensionalReduction} where the entanglement entropy was computed using dimensional reduction. 

\subsection{Multiplet structure for some initial states}
We begin by expressing our chosen initial states in a form which is amenable to the evolution with the QPP.  For simplicity, though, we specialize to the case where there are two fermions per unit cell. An analogous treatment can be performed, albeit with tedium, for ${\mathtt{n}}>2$, see Appendix~\ref{sec:AppA}. With this choice our initial state is
\beqa
    \ket{\psi} &=& \prod_{\boldsymbol{j}=0}^{\boldsymbol{L}/\boldsymbol{\mu}-1}\left(\sum_{\boldsymbol{m}=0}^{\boldsymbol{\mu}-1} a^{(1)}_{\boldsymbol{m}}c^\dagger_{\boldsymbol{\mu}\boldsymbol{j}-\boldsymbol{m}}\right)\left(\sum_{\boldsymbol{m}'=0}^{\boldsymbol{\mu}-1} a^{(2)}_{\boldsymbol{m}'}c^\dagger_{\boldsymbol{\mu}\boldsymbol{j}-\boldsymbol{m}'}\right) \ket{0},
\eeqa
where $\boldsymbol{L} = (L_x,L_y)$, $\boldsymbol{\mu}= (\mu_x,\mu_y)$ and $\boldsymbol{L}/\boldsymbol{\mu} = (L_x/
\mu_x,L_y/\mu_y)$ since we are considering two spatial dimensions. 
We saw in the one dimensional case that, in order to employ the QPP approach to dynamics, we were required to split the system into a set of hydrodynamic fluid cells. The increased dimensionality of the system means that there are numerous choices for this splitting. 
However, because of the specific geometry of both our system and subsystem,  a natural choice is to take the fluid cells to be of length $\Delta$  in the $x$-direction but cover the $y$-direction entirely, i.e. $\boldsymbol{\Delta}=(\Delta,L_y)$. Moreover, the  fluid cell should contain a large number of lattice sites in the $x$ direction but also be small on the scale of the subsystem $1\ll \Delta \ll \ell$. There is no such requirement in the $y$-direction.

As in the previous section, we split the $x$ coordinate into $x=x_0+\varkappa$ with $x_0\in[1,L/\Delta]$ and $\varkappa\in[0,\Delta-1]$ but leave the $y$ coordinate as it is.  We also introduce the partially Fourier transformed creation operators inside the fluid cell
\begin{eqnarray}
b^\dag_{x_0,\boldsymbol{q}}=\frac{1}{\sqrt{\Delta L_y}}\sum_{y=1}^{L_y}\sum_{\varkappa=0}^{\Delta-1} e^{-iq_x\varkappa-iq_yy}c^\dag_{x_0+\varkappa,y},
\end{eqnarray}
which will serve as our quasiparticle creation operators for the QPP.  Using these, the initial state can then be expressed as~\cite{molly},
   \beqa 
   \ket{\psi} =\prod_{x_0=0}^{L/\Delta-1} \prod_{\boldsymbol{p}\in \frac{2\pi}{\boldsymbol{\Delta}}\mathbb{Z}^2_{\boldsymbol{\Delta}/\boldsymbol{\mu}}}\sum_{\vk \in \frac{2\pi}{\boldsymbol\mu}\mathbb{Z}^2_{\boldsymbol{\mu}} }\sum_{\vk' \in \frac{2\pi}{\boldsymbol\mu}\mathbb{Z}^2_{\boldsymbol{\mu}} }f^{(1)}_{\boldsymbol{p}+\vk}f^{(2)}_{\boldsymbol{p}+\vk'} b^\dagger_{x_0,\boldsymbol{p}+\vk} b^\dagger_{x_0,\boldsymbol{p}+\vk'}\ket{0}.
\eeqa
Here the fluid cell momenta are quantized according to  $\boldsymbol{q}\in \frac{2\pi}{\boldsymbol{\Delta}}\mathbb{Z}^2_{\boldsymbol{\Delta}}$ where we use the notation $\frac{2\pi}{\boldsymbol{a}}\mathbb{Z}^n_{\boldsymbol{b}}=\frac{2\pi}{a_1}\mathbb{Z}_{b_1}\otimes\frac{2\pi}{a_2}\mathbb{Z}_{b_2}\otimes\dots\otimes \frac{2\pi}{a_n}\mathbb{Z}_{b_n}$ for some vectors $\boldsymbol{a},\boldsymbol{b}$. This momentum has been split into $\boldsymbol{q}=\boldsymbol{p}+\boldsymbol{k}$ with $\boldsymbol{p}\in \frac{2\pi}{\boldsymbol{\Delta}}\mathbb{Z}^2_{\boldsymbol{\Delta}/\boldsymbol{\mu}}$ and $\boldsymbol{k}\in \frac{2\pi}{\boldsymbol{\mu}}\mathbb{Z}^2_{\boldsymbol{\mu}}$ thereby reflecting the symmetry of the unit cell. The functions $f^{(\lambda)}_{\boldsymbol{p}+\vk}$ for  $\lambda=1,2$ encode the specific coefficients which identify the state and are given by
\begin{equation}
f^{(\lambda)}_{\boldsymbol{p}+\vk}=\frac{1}{\sqrt{\mu_x\mu_y}} \sum_{\boldsymbol{m}=0}^{\boldsymbol{\mu}-1}a^{(\lambda)}_{\boldsymbol{m}}e^{-i(\boldsymbol{p}+\vk)\cdot \boldsymbol{m}}.
\end{equation}
As with the symmetry preserving state in one dimension~\eqref{eq:symm_preserve_state_1d}, this decomposition is exact, owing to the initial state being a product state. We see, however, that this state has a much more complicated quasiparticle structure than the pair structure seen in \eqref{eq:symm_preserve_state_1d}. In the previous case, the initial state contained pairs of correlated quasiparticles in which the members of a pair have momenta $p$ or $-p$. In the present case, however,  there are instead correlated multiplets of $\mu_x\mu_y$ quasiparticles, identified with a value of $\boldsymbol{p}$ with each member of the multiplet taking a value of momentum in $\boldsymbol{p}+\vk$ where $\vk \in \frac{2\pi}{\boldsymbol\mu}\mathbb{Z}_{\boldsymbol{\mu}}$. For $\mu_x=2,\mu_y=1$ we have a state which has single site shift symmetry in the $y$-direction and two site shift symmetry in $x$ which recovers the pair structure of the one dimensional case seen previously. In that instance the system, in effect, consists of $L_y$ decoupled one dimensional systems and the analysis simplifies considerably, mirroring the previous section. 

These complications aside, the initial state is a simple product over fluid cells and then values of $\boldsymbol{p}$ which are quantized outside the unit cell identified by the shift symmetry,  $\boldsymbol{\mu}$. The same is true also for the density matrix which we can express as 
\begin{equation}
    \rho(0) = \prod_{x_0=1}^{L/\Delta}\prod_{\boldsymbol{p}\in \frac{2\pi}{\boldsymbol{\Delta}}\mathbb{Z}^2_{\boldsymbol{\Delta}/\boldsymbol{\mu}}} \rho_{x_0,\boldsymbol{p}}
\label{eq:rhoinitialstate}
\end{equation}
where $\rho_{x_0,\boldsymbol{p}}$ is the density matrix describing a correlated multiplet of excitations and has the form
\beqa\nonumber
    \rho_{x_0,\boldsymbol{p}} &=& \hspace{-10pt}\sum_{\vk,\vk' \in \frac{2\pi}{\boldsymbol\mu}\mathbb{Z}^2_{\boldsymbol{\mu}}} \hspace{-10pt}f^{(1)}_{\boldsymbol{p}+\vk}f^{(2)}_{\boldsymbol{p}+\vk'} b^\dagger_{x_0,\boldsymbol{p}+\vk} b^\dagger_{x_0,\boldsymbol{p}+\vk'} \ket{0} \bra{0}_{x_0}\hspace{-10pt}\sum_{\vk'',\vk''' \in \frac{2\pi}{\boldsymbol\mu}\mathbb{Z}^2_{\boldsymbol{\mu}}} \hspace{-10pt}{f^{(1)}}^*_{\hspace{-10pt}\boldsymbol{p}+\vk''}{f^{(2)}}^*_{\hspace{-10pt}\boldsymbol{p}+\vk'''} b_{x_0,\boldsymbol{p}+\vk''} b_{x_0,\boldsymbol{p}+\vk'''} \\\nonumber
    &=& \hspace{-10pt}\sum_{\vk,\vk' \in \frac{2\pi}{\boldsymbol\mu}\mathbb{Z}^2_{\boldsymbol{\mu}}} \hspace{-10pt}f^{(1)}_{\boldsymbol{p}+\vk}f^{(2)}_{\boldsymbol{p}+\vk'} b^\dagger_{x_0,\boldsymbol{p}+\vk} b^\dagger_{x_0,\boldsymbol{p}+\vk'}\Big( \!\!\prod_{\boldsymbol{q} \in \frac{2\pi}{\boldsymbol\mu}\mathbb{Z}^2_{\boldsymbol{\mu}}} (1-\hat{n}_{x_0,\boldsymbol{p}+\boldsymbol{q}})\Big)\\
    &&\hspace{80pt}\times \sum_{\vk'',\vk''' \in \frac{2\pi}{\boldsymbol\mu}\mathbb{Z}_{\boldsymbol{\mu}}} \hspace{-10pt}{f^{(1)}}^*_{\hspace{-10pt}\boldsymbol{p}+\vk''}{f^{(2)}}^*_{\hspace{-10pt}\boldsymbol{p}+\vk'''} b_{x_0,\boldsymbol{p}+\vk''} b_{x_0,\boldsymbol{p}+\vk'''}
\eeqa
where in the first line we have introduced the fluid cell vacuum defined via $ \ket{0} \bra{0} = \prod_{x_0=0}^{L/\Delta-1}  \ket{0}\bra{0}_{x_0}$ and in the second line we have used the representation of the vacuum density matrix,
\begin{eqnarray}
    \ket{0} \bra{0} = \prod_{x_0=0}^{L/\Delta-1}  \ket{0}\bra{0}_{x_0}=\prod_{x_0=0}^{L/\Delta-1} \prod_{\boldsymbol{p}\in \frac{2\pi}{\boldsymbol{\Delta}}\mathbb{Z}^2_{\boldsymbol{\Delta}/\boldsymbol{\mu}}}\prod_{\vk \in \frac{2\pi}{\boldsymbol\mu}\mathbb{Z}^2_{\boldsymbol{\mu}} }(1-\hat{n}_{x_0,\boldsymbol{p}+\boldsymbol{k}}).
\end{eqnarray}  
Since, in the end we will be interested in tracing out over certain quasiparticles  we can already separate out those terms which survive this procedure. 
 These contain the same type of creation operators on the left as the annihilation operators on the right. In this specific case there are two choices, differing by a sign, which give
\begin{eqnarray}\nonumber
    b^\dagger_{x_0,\boldsymbol{p}+\vk} b^\dagger_{x_0,\boldsymbol{p}+\vk'} \ket{0} \bra{0}_{x_0}b_{x_0,\boldsymbol{p}+\vk''} b_{x_0,\boldsymbol{p}+\vk'''}
= \left(\delta_{\vk,\vk'''}\delta_{\vk',\vk''}-\delta_{\vk,\vk''}\delta_{\vk',\vk'''}\right)\hat{n}_{x_0,\boldsymbol{p}+\vk}\hat{n}_{x_0,\boldsymbol{p}+\vk'}\\
\times\prod_{\boldsymbol{q}\neq\vk,\vk' }(1-\hat{n}_{x_0,\boldsymbol{p}+\boldsymbol{q}}) + \dots\quad
\end{eqnarray}
 where the dots are the other terms which have to disappear when tracing out. Substituting in the above this gives
\beqa\nonumber
    \rho_{x_0,\boldsymbol{p}} &=& \hspace{-10pt}\sum_{\vk,\vk'\in \frac{2\pi}{\boldsymbol\mu}\mathbb{Z}^2_{\boldsymbol{\mu}}} \hspace{-10pt}\left[f^{(1)}_{\boldsymbol{p}+\vk}{f^{(1)}}^*_{\hspace{-10pt}\boldsymbol{p}+\vk'}f^{(2)}_{\boldsymbol{p}+\vk'}{f^{(2)}}^*_{\hspace{-10pt}\boldsymbol{p}+\vk} - |f^{(1)}_{\boldsymbol{p}+\vk}|^2|f^{(2)}_{\boldsymbol{p}+\vk}|^2\right]\hat{n}_{x_0,\boldsymbol{p}+\vk}\hat{n}_{x_0,\boldsymbol{p}+\vk'}\\
    && \hspace{80pt}\times\prod_{\boldsymbol{q}\neq\vk,\vk' }(1-\hat{n}_{x_0,\boldsymbol{p}+\boldsymbol{q}}) +\dots\\
    &=& \hspace{-10pt} \sum_{\vk,\vk'\in \frac{2\pi}{\boldsymbol\mu}\mathbb{Z}^2_{\boldsymbol{\mu}}} F_{k_x,k_y}^{k_x',k_y'}(\boldsymbol{p})\,\hat{n}_{x_0,\boldsymbol{p}+\vk}\hat{n}_{x_0,\boldsymbol{p}+\vk'}\hspace{-10pt}\prod_{\boldsymbol{q}\neq\vk,\vk' }(1-\hat{n}_{x_0,\boldsymbol{p}+\boldsymbol{q}}) + \dots \hspace{0.2cm}. 
    \label{eq:rhopfull}
\eeqa
where in the second line we have defined the quantity 
\begin{eqnarray}
    F_{k_x,k_y}^{k_x',k_y'}(\boldsymbol{p})=f^{(1)}_{\boldsymbol{p}+\vk}{f^{(1)}}^*_{\hspace{-10pt}\boldsymbol{p}+\vk'}f^{(2)}_{\boldsymbol{p}+\vk'}{f^{(2)}}^*_{\hspace{-10pt}\boldsymbol{p}+\vk} - |f^{(1)}_{\boldsymbol{p}+\vk}|^2|f^{(2)}_{\boldsymbol{p}+\vk}|^2.
\end{eqnarray} Note that this function vanishes for $\vk=\vk'$ and so the state contains two quasiparticles which are distinct members of the multiplet. All other members of the multiplet are projected out by the term $\prod_{\boldsymbol{q}\neq\vk,\vk' }(1-\hat{n}_{x_0,\boldsymbol{p}+\boldsymbol{q}})$ ensuring that there are only two fermions per unit cell. 

\subsection{Quasiparticle picture dynamics}
We are now in a position to study the time evolution of our system using the QPP. As in the one dimensional case, we approximate the actual time evolution of the fluid cell creation operators in a semiclassical fashion by propagating them to another fluid cell in a way which depends their velocity. 
Thus, the specific evolution of quasiparticle pairs is determined by the selection of fluid cells, which, in turn, is shaped by the geometry of the subsystem under consideration.
In the one dimensional case, there is a unique choice of fluid cells however this is not so in higher dimensions.  For our choice of $A$, since the fluid cells entirely cover the $y$ direction and $L_y$ may be finite the evolution is given by
\begin{equation}
\label{eq:semiclassical_evolution}  e^{iHt}b^\dag_{x_0,\boldsymbol{p}+\vk} e^{-iHt} \approx b^\dag_{x_t(\boldsymbol{p}+\vk),\boldsymbol{p}+\vk},
\end{equation}
where now $x_t(\boldsymbol{p}+\vk)=x_0+v_x(\boldsymbol{p}+\vk)t.$ Therefore, the dynamics of the operators depends only on the $x$ component of velocity. Once again there are exceptional cases where a Bogoliubov rotation is necessary to capture the correct behavior~\cite{DimensionalReduction}. We postpone a discussion of these states till later and assume that our states do not fall into this category. Upon using~\eqref{eq:semiclassical_evolution} in the expression~\eqref{eq:rhopfull}, we have that 
\begin{eqnarray}
    \rho(t) &=&\prod_{\boldsymbol{p}\in \frac{2\pi}{\boldsymbol{\Delta}}\mathbb{Z}^2_{\boldsymbol{\Delta}/\boldsymbol{\mu}}} \prod_{x_0=1}^{L/\Delta}\rho_{x_0,\boldsymbol{p}}(t),\\\nonumber
  \rho_{x_0,\boldsymbol{p}}(t)  &=&\hspace{-10pt} \sum_{\vk,\vk'\in \frac{2\pi}{\boldsymbol\mu}\mathbb{Z}^2_{\boldsymbol{\mu}}} F_{k_x,k_y}^{k_x',k_y'}(\boldsymbol{p})\,\hat{n}_{x_t(\boldsymbol{p}+\vk),\boldsymbol{p}+\vk}\,\hat{n}_{x_t(\boldsymbol{p}+\vk),\boldsymbol{p}+\vk'}\prod_{\boldsymbol{q}\neq\vk,\vk' }(1-\hat{n}_{x_t(\boldsymbol{p}+\vk),\boldsymbol{p}+\boldsymbol{q}}) + \dots \,, 
\end{eqnarray}
where again the ellipsis denotes terms which will not survive the trace. 

The multiplet structure of the initial state means that determining which quasiparticles should be traced out and when, becomes more involved. To simplify matters, we make the further restriction to the case where we have two site translational symmetry along both the $x$ and $y$ directions. In that case, the multiplet contains 4 quasiparticles, distinguished by the values of $k_{x,y}=0,\pi$. For the tight binding chain this translates to quasiparticle velocities of the multiplet being $\boldsymbol{v}(\boldsymbol{p}+\vk)=(\pm\sin(p_x),\pm\sin(p_y))$.  We can now proceed as before and determine the reduced density matrix for the multiplet $\rho_{A,x_0,\boldsymbol{p}}(t)$ when some members are no longer present inside $A$. Let us suppose that after a time $t$, the quasiparticle labeled by $\vk=(\pi,\pi)$ is no longer inside $A$, i.e. $x_t(\boldsymbol{p}+\boldsymbol{\pi})=x_0-\sin(p_x)t\notin A$, which also implies that the quasiparticle with $\vk=(\pi,0)$ is also not inside $A$. Performing the trace over these particles we arrive, after some manipulations, at  
\begin{eqnarray}\nonumber
\rho_{A,x_0,\boldsymbol{p}}(t)&=&\left[n(\boldsymbol{p})\hat{n}_{x_t(\boldsymbol{p}),\boldsymbol{p}}+(1-n(\boldsymbol{p}))(1-\hat{n}_{x_t(\boldsymbol{p}),\boldsymbol{p}})\right]\\ \nonumber
&&\hspace{80pt}\times\left[n(\bar{\boldsymbol{p}})\hat{n}_{x_t(\bar{\boldsymbol{p}}),\bar{\boldsymbol{p}}}+(1-n(\bar{\boldsymbol{p}}))(1-\hat{n}_{x_t(\bar{\boldsymbol{p}}),\bar{\boldsymbol{p}}})\right]\\
&=&\frac{1}{1+e^{-\eta(\boldsymbol{p})}}\frac{1}{1+e^{-\eta(\bar{\boldsymbol{p}})}}e^{-\eta(\boldsymbol{p})\hat{n}_{x_t({\boldsymbol{p})},\boldsymbol{p}}+\eta(\bar{\boldsymbol{p}})\hat{n}_{x_t(\bar{\boldsymbol{p}}),\bar{\boldsymbol{p}}}}\ .
\end{eqnarray}
Here, we have introduced the shorthand notation $\bar{\boldsymbol{p}}=\boldsymbol{p}+(0,\pi)$ and used the functions $\eta(\boldsymbol{p})$ which are defined as in~\eqref{eq:eta}. The presence of the fermion occupation functions $n(\boldsymbol{p})$ can be derived by carefully considering the form of the function  $F_{k_x,k_y}^{k_x',k_y'}(\boldsymbol{p})$ which we detail in  Appendix~\ref{sec:AppA}. Carrying out the same procedure assuming, instead, that it is the $\vk=(0,\pi)$ and $\vk=(0,0)$ quasiparticle which do not remain in the subsystem we arrive at a similar form,
\begin{eqnarray}\nonumber
\rho_{A,x_0,\boldsymbol{p}_\pi}(t)\!\!\!&=&\!\!\!\left[n(\boldsymbol{p}_\pi)\hat{n}_{x_t(\boldsymbol{p}_\pi),\boldsymbol{p_\pi}}\!\!+(1-n(\boldsymbol{p}_\pi))(1-\hat{n}_{x_t(\boldsymbol{p}_\pi),\boldsymbol{p}_\pi})\right]\\\nonumber
&& \hspace{80pt}\times
\left[n(\bar{\boldsymbol{p}}_\pi)\hat{n}_{x_t(\bar{\boldsymbol{p}}_\pi),\bar{\boldsymbol{p}}_\pi}\!\!+(1-n(\bar{\boldsymbol{p}}_\pi))(1-\hat{n}_{x_t(\bar{\boldsymbol{p}}_\pi),\bar{\boldsymbol{p}}_\pi})\right]\\
&=&\frac{1}{1+e^{-\eta(\boldsymbol{p})}}\frac{1}{1+e^{-\eta(\bar{\boldsymbol{p}}_\pi)}}e^{-\eta(\boldsymbol{p}_\pi)\hat{n}_{x_t({\boldsymbol{p})},\boldsymbol{p}_\pi}-\eta(\bar{\boldsymbol{p}}_\pi)\hat{n}_{x_t(\bar{\boldsymbol{p}}_\pi),\bar{\boldsymbol{p}}_\pi}},
\end{eqnarray}
where we have used the notation $\boldsymbol{p}_\pi=\boldsymbol{p}+(\pi,0)$ and $\bar{\boldsymbol{p}}_\pi=\boldsymbol{p}+(\pi,\pi)$. 

We are now naturally led to the same decomposition of the reduced density matrix in terms of mixed and pure terms as we found in the previous section~\eqref{eq:decomposition}, i.e. 
\begin{eqnarray}\label{eq:decomposition_2d}
    \rho_A(t)\approx \rho_{\rm mixed}(t)\otimes \rho_{\rm pure}(t)
\end{eqnarray}
but with
\begin{eqnarray}\label{eq:pure_d2}
    \rho_{\rm pure}(t)&=&\prod_{\boldsymbol{p}}\prod_{\{x_0|x_t(\boldsymbol{p})\,\& \, x_t(\boldsymbol{p}_\pi)\in A\}}\rho_{x_0,\boldsymbol{p}}(t)\\\label{eq:mixed_2d}
    \rho_{\rm mixed}(t)&=&\prod_{\boldsymbol{p}}\prod_{\{x_0|x_t(\boldsymbol{p})\in A, x_t(\boldsymbol{p}_\pi)\notin A\}}\!\!\!\rho_{A,x_0,\boldsymbol{p}}(t)\prod_{\{x_0|x_t(\boldsymbol{p})\notin A, x_t(\boldsymbol{p}_\pi)\in A\}}\!\!\!\rho_{A,x_0,\boldsymbol{p}_\pi}(t).
\end{eqnarray}
The reasoning is similar to the one dimensional case:  for each $x_0$ and $\boldsymbol{p}$ we check which members of the multiplet are inside $A$.  If all are present, they contribute to $\rho_{\rm pure}$,  if only two members are inside $A$ then they contribute instead to $\rho_{\rm mixed}$,  while if none are present they do not contribute to $\rho_A(t)$. 

The mixed part can subsequently be expressed in terms of the entanglement Hamiltonian  as follows
\begin{eqnarray}
   \rho_{\rm mixed}(t)&=&\frac{1}{\mathcal{Z}_A}\exp{-K_{A,QP}(t)}\\\nonumber
   K_{A,QP}(t)&=&\sum_{\boldsymbol{p}}\sum_{\{x_0|x_t(\boldsymbol{p})\in A, x_t(\boldsymbol{p}_\pi)\notin A\}}\hspace{-30pt}\eta(\boldsymbol{p})\hat{n}_{x_t({\boldsymbol{p})},\boldsymbol{p}}+\eta(\bar{\boldsymbol{p}})\hat{n}_{x_t(\bar{\boldsymbol{p}}),\bar{\boldsymbol{p}}}~,\\
   &&+\sum_{\boldsymbol{p}}\sum_{\{x_0|x_t(\boldsymbol{p})\notin A, x_t(\boldsymbol{p}_\pi)\in A\}}\hspace{-30pt}\eta(\boldsymbol{p}_\pi)\hat{n}_{x_t({\boldsymbol{p})},\boldsymbol{p}_\pi}+\eta(\bar{\boldsymbol{p}}_\pi)\hat{n}_{x_t(\bar{\boldsymbol{p}}_\pi),\bar{\boldsymbol{p}}_\pi}.
\end{eqnarray}
Note that the sum over $\boldsymbol{p}$ is still over the restricted set of values $\frac{2\pi}{\boldsymbol{\Delta}}\mathbb{Z}^2_{\boldsymbol{\Delta}/\boldsymbol{\mu}}$. However this can be combined with the various $\overline{\boldsymbol{p}}$ and $\boldsymbol{p}_\pi$ appearing to give a single sum over unrestricted momenta:
\begin{equation}
    K_{A,QP}(t)=\sum_{p_x\in \frac{2\pi}{\Delta}\mathbb{Z}_\Delta}  \sum_{p_y \in \frac{2\pi}{L_y}\mathbb{Z}_{L_y}} \sum_{\{x_0|x_t(\boldsymbol{p})\notin A, x_t(\boldsymbol{p}_\pi)\in A\}} \eta(\boldsymbol{p}) \hat{n}_{x_t(\boldsymbol{p}),\boldsymbol{p}}.
\end{equation}
Once again we should convert this back to real space. In doing so we pass from sums over lattice sites in the $x$ direction to integrals but leave the $y$-direction as discrete. From this we obtain 
\beqa
     K_{A,QP}{(t)} &=& \!\!\!\sum_{y,y'} \int_0^l dx \int_{0< x-z\leq\ell} \hspace{-30pt}dz\,[ \mathcal{K}^{(y,y')}_L(x,z;t)+ \mathcal{K}^{(y,y')}_R(x,z;t)]c_{x,y}^\dagger c_{x-z,y'} \\
     &=&\!\!\!\sum_{y} \sum_{z_y} \int_0^l \!dx \!\!\int_{0< x-z_x\leq\ell}\hspace{-35pt} dz_x\,[ \mathcal{K}^{(z_y)}_L(x,z_x;t)+ \mathcal{K}^{(z_y)}_R(x,z_x;t)]c_{x,y}^\dagger c_{x-z_x,y-z_y}
     \label{eq:finalEH}
\eeqa
where $z_y =y'-y $ and  
\begin{eqnarray}\label{eq:realspacecoeff_L}
    \mathcal{K}^{(z_y)}_{L}(x,z_x;t) \!\!&=& \!\!\!\!\frac{1}{L_y}\sum_{q_y}e^{-iq_yz_y}\!\int_{q_x<0}\!\frac{dq_x}{2\pi}\eta(\boldsymbol{q}) \Theta(\max(l+2v_x(q_x)t,0)+x)e^{-iq_x z_x}\,\,\,\\\label{eq:realspacecoeff_R}
     \mathcal{K}^{(z_y)}_{R}(x,z_x;t) \!\!&=& \!\!\!\!\frac{1}{L_y}\sum_{q_y}e^{-iq_yz_y}\!\int_{q_x>0}\!\frac{dq_x}{2\pi}\eta(\boldsymbol{q}) \Theta(\min(2v_x(q_x)t,l)-x)e^{-iq_x z_x}
\end{eqnarray}
Remarkably, the result is essentially the same as in 1D with an extra Fourier transform. We have two terms originating from the left and right boundaries of the subsystem, each of which displays a light cone structure centered around these surfaces.  
Although we have focused on two-dimensional systems, it is straightforward to extend the discussion to higher dimensions, obtaining a very similar result. Specifically, we will show in appendices \ref{sec:AppA} and \ref{sec:AppC} that an entanglement Hamiltonian of the form \eqref{eq:finalEH} emerges in all situations in which the reduced density matrix equilibrates to a GGE in the long time limit, that is when
\begin{equation}
    \lim_{t\to \infty} \rho_A(t) = \rho_\infty = \rho_{GGE}.
\end{equation}
A counterexample to this condition will be provided in section \ref{sec:crossed}.
\subsection{Analytic checks}

We can once again perform some analytic checks on our results for $\rho_A(t)$ and in particular $K_{A,QP}(t)$ by using it to calculate certain quantities and comparing to known results. 

For the entanglement entropy we can use the decomposition~\eqref{eq:decomposition_2d},\eqref{eq:pure_d2} and~\eqref{eq:mixed_2d} to find
\begin{eqnarray}
S_A(t)&=&-\sum_{\boldsymbol{p}}\sum_{{\{x_0|x_t(\boldsymbol{p})\in A, x_t(\boldsymbol{p}_\pi)\notin A\}}}\operatorname{Tr}[\rho_{A,x_0,\boldsymbol{p}}(t)\log \rho_{A,x_0,\boldsymbol{p}}(t)]\\
   &&-\sum_{\boldsymbol{p}}\sum_{{\{x_0|x_t(\boldsymbol{p})\notin A, x_t(\boldsymbol{p}_\pi)\in A\}}}\operatorname{Tr}[\rho_{A,x_0,\boldsymbol{p}_\pi}(t)\log \rho_{A,x_0,\boldsymbol{p}_\pi}(t)]
\end{eqnarray}
where only the mixed part contributes. In the thermodynamic limit this becomes
\begin{eqnarray}
    S_A(t)= \sum_{p_y}\int \frac{dp_x}{2\pi}\min(2|v_x(p_x)|t,l)[n(\boldsymbol{p})\log n(\boldsymbol{p})+(1-n(\boldsymbol{p}))\log (1-n(\boldsymbol{p}))]
   \label{eq:dimredentropy}
\end{eqnarray}
which is in agreement with known results~\cite{DimensionalReduction,molly}. The extension to arbitrary R\'enyi entropies is immediate and is again consistent with the findings of \cite{DimensionalReduction}.

The full counting statistics of the charge can similarly be calculated. For this we find 
\begin{eqnarray}\nonumber
\log Z_A(\alpha,t)&=&\sum_{\boldsymbol{p}}\sum_{{\{x_0|x_t(\boldsymbol{p})\in A, x_t(\boldsymbol{p}_\pi)\in A\}}}\log\operatorname{Tr}[e^{\alpha[\hat{n}_{x_t}(\boldsymbol{p})+\hat{n}_{x_t}(\bar{\boldsymbol{p}})+\hat{n}_{x_t}(\boldsymbol{p}_\pi)+\hat{n}_{x_t}(\bar{\boldsymbol{p}}_\pi)]}\rho_{x_0,\boldsymbol{p}}(t)]\\\nonumber
&&+\sum_{\boldsymbol{p}}\sum_{{\{x_0|x_t(\boldsymbol{p})\in A, x_t(\boldsymbol{p}_\pi)\notin A\}}}\log\operatorname{Tr}[e^{\alpha[\hat{n}_{x_t}(\boldsymbol{p})+\hat{n}_{x_t}(\bar{\boldsymbol{p}})]}\rho_{A,x_0,\boldsymbol{p}}(t)]\\
   &&+\sum_{\boldsymbol{p}}\sum_{{\{x_0|x_t(\boldsymbol{p})\notin A, x_t(\boldsymbol{p}_\pi)\in A\}}}\log\operatorname{Tr}[e^{\alpha[\hat{n}_{x_t}(\boldsymbol{p}_\pi)+\hat{n}_{x_t}(\bar{\boldsymbol{p}}_\pi)]}\rho_{A,x_0,\boldsymbol{p}_\pi}(t)]
\end{eqnarray}
where the first line comes from the pure part and the second and third from the mixed. In the thermodynamic limit this can be evaluated to give
\begin{eqnarray}\nonumber
    \log Z_A(\alpha,t)&=& \sum_{k_y \in \frac{2\pi}{L_y}\mathbb{Z}_{L_y}}\int \frac{dk_x}{2\pi}\min(2|v_x(k_x)|t,\ell) \log(1-n(\vk)+n(\vk)e^{-\alpha})  \\\label{eq:FCS_2d}
    &&+ \int_{k>0} \frac{dk_x}{2\pi}  \left[\ell-\min(2|v_x(k_x)|t,\ell)\right] \alpha L_y,
\end{eqnarray}
Here, the first term originates from the mixed component, while the second arises from the pure component. 
The latter takes this specific form because we are considering symmetry-preserving states, which are charge eigenvalues. 
Although Eq. \eqref{eq:FCS_2d} could have been derived using the methods of Ref. \cite{DimensionalReduction}, it appears that, to the best of our knowledge, this explicit calculation has not been performed before.

\section{Examples and Numerical checks}
\label{sec:examples}
In this section, we investigate the entanglement Hamiltonian for some specific examples and make a comparison between the analytic prediction~\eqref{eq:finalEH} and exact numerics.  The numerical procedure is straightforward and relies upon the relation between the correlation matrix and the entanglement Hamiltonian~\eqref{eq:peschel}. We start from the exact expression for $C_A(t)$, which we arrange in an $\ell L_y \times \ell L_y$ matrix. This can be numerically diagonalized in order to find its eigenvalues $\lambda_i(t)$, which lie between 0 and 1, and its eigenvectors $\boldsymbol{V}_i(t)$. 
Having found these, the entanglement Hamiltonian can be obtained by inverting \eqref{eq:peschelformula},
\begin{equation}
    K_{A,exact}^{(t)} = V^T(t) \cdot \operatorname{diag}(\log(\lambda^{-1}_i(t)-1)) \cdot V(t)
    \label{eq:nunmericalexact}
\end{equation}
where $V(t)$ is the matrix whose columns are the eigenvectors $\boldsymbol{V}_i(t)$. This scheme is exact, however we wish to use to compare to our results for $K_{A,QP}(t)$ which are derived using the decomposition of~\eqref{eq:decomposition_2d}. In our expression some parts of the exact reduced density matrix, which is entirely mixed at finite time, have been
approximated as being part of $\rho_{\rm pure}(t)$. 
In order to properly compare~\eqref{eq:finalEH} with the numerical result,
we should remove from \eqref{eq:nunmericalexact} the eigenvalues which are too close to $0$ and $1$ which form a subspace that is approximated $\rho_{\rm pure}(t)$.  This is done by introducing a cutoff $\epsilon$ in the spectrum. The eigenspace pertaining to eigenvalues which are in the range $[0,\epsilon]$ and $[1-\epsilon,1]$ is omitted from~\eqref{eq:nunmericalexact} and the resulting $K_{A,\rm{approx}}(t)$ compared to $K_{A,QP}(t)$.  Since the distribution of the eigenvalues of the correlation matrix is strongly dependent on the choice of initial state, it is natural that also the cutoff will be. We take taking as reference examples the states considered in \cite{DimensionalReduction}; the collinear, staggered and diagonal dimers and use a cutoff $\epsilon=10^{-4}$ for the first two and $\epsilon = 10^{-6}$ for the last.

The procedure is more efficient if a compact closed form expression for the correlation matrix elements can be found. In appendix \ref{sec:AppB} we derive exact analytical expressions for these at arbitrary times, in a regime in which we take a thermodynamic limit only along the $x$ axis.

\subsection{Collinear dimer}
\begin{figure}[h]
    \centering
    \includegraphics[width=0.8\linewidth]{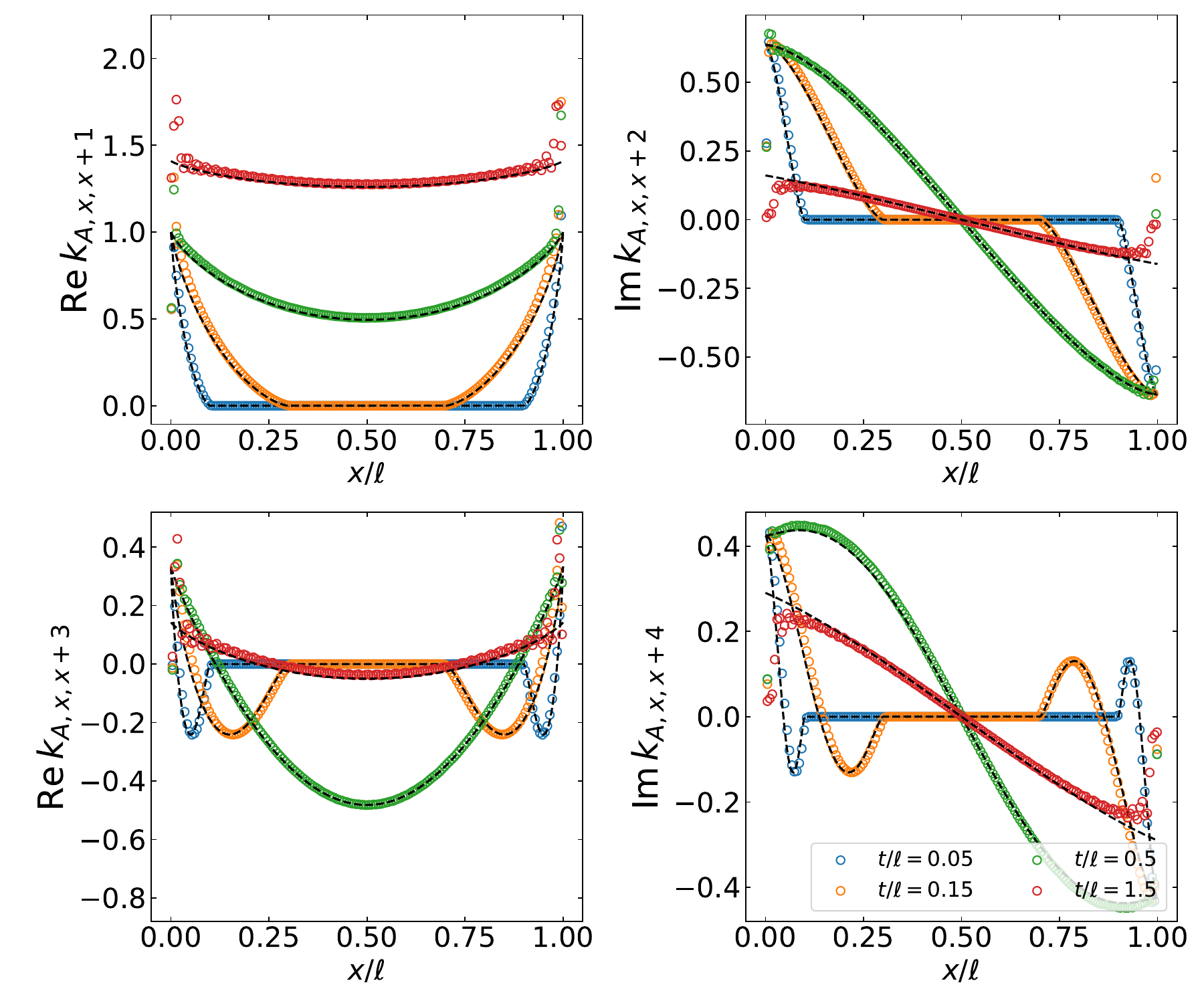}
    \caption{Quench from the collinear dimer state with $L_x=800$ and $L_y=4$. The dashed lines are the quasiparticle predictions for the term $\mathcal{K}^{(z_y)}_{R}(x,z_x;t) + \mathcal{K}^{(z_y)}_{L}(x,z_x;t) $ , plotted for $z_y=0$ (since all other values give zero), and values of $z_x$ ranging from 1 to 4. The symbols are exact results obtained combining the exact expressions for the correlation matrix and equation \eqref{eq:peschelformula}. The agreement is excellent except for some slight deviations at the edges of the interval, where finite size effects appear. }
    \label{fig:colldim}
\end{figure}

As a first example we take the collinear dimer state. This is defined as
\begin{equation}
    \ket{\text{CD}} = \prod_{x=1}^{L_x/2}\prod_{y=1}^{L_y} \frac{1}{\sqrt{2}}\left(c^\dagger_{2x,y} -c^\dagger_{2x+1,y}\right) \ket{0}.
\end{equation}
As shown in the appendix, this situation is almost trivial since the correlation matrix is block diagonal with respect to y and each block corresponds to the 1-dimensional dimer state in the $x$-direction, which is essentially the same as the one already considered in \cite{1D} \footnote{There are some minor sign differences, which arise from the fact that in \cite{1D} the state considered was actually the one with occupation functions $n(k) = \frac{1}{2}(1+\cos(k_x))$ instead of $\frac{1}{2}(1-\cos(k_x))$. At the quasiparticle level this just accounts for a total minus in front on the entanglement Hamiltonian: this can be seen from the fact that the plots \ref{fig:colldim} are specular to the ones of \cite{1D}.}
\begin{equation}
    C_{\boldsymbol{x},\boldsymbol{x}'}(t)= \delta_{y,y'}C^{1D}_{x,x'}(t)
\end{equation}
where the 1D correlation matrix is 
\begin{equation}\label{eq:cd_correl}
     C^{1D}_{x,x'}(t) = C^{\infty}-e^{-i\pi/2 (x+x')}\frac{i(x-x')}{4t}J_{x-x'}(2t)
\end{equation}
with $C^{\infty}=\frac{1}{2}\left(\delta_{x,x'} - \frac{1}{2}(\delta_{x,x'+1}+\delta_{x,x'-1})\right)$ the correlation matrix at $t=\infty$.
This implies that  if $z_y\neq 0$ the elements of the entanglement Hamiltonian will be zero, while if $z_y=0$ the results should be identical to the 1D dimer state. This is immediately confirmed by the quasiparticle prediction.  From \eqref{eq:cd_correl} we find that the occupation functions only depend on $k_x$,
\begin{equation}
    n_{CD}(\vk)= \frac{1}{2}(1-\cos(k_x)) = n_{1D}(k_x) \rightarrow \eta_{CD}(k_x) = 2 \log \cot\frac{k_x}{2}
\end{equation}therefore $\mathcal{K}^{(k_y)}_{L,R} = \mathcal{K}_{L,R}$ (with no $k_y$ dependence) and hence $\mathcal{K}^{(z_y)}_{L,R} =\delta_{z_y,0}\mathcal{K}_{L,R}$.
This identity is indeed confirmed by numerical calculations, as shown in figure \ref{fig:colldim}.

\subsection{Staggered dimer}
\begin{figure}[h]
    \centering
    \includegraphics[width=0.8\textwidth]{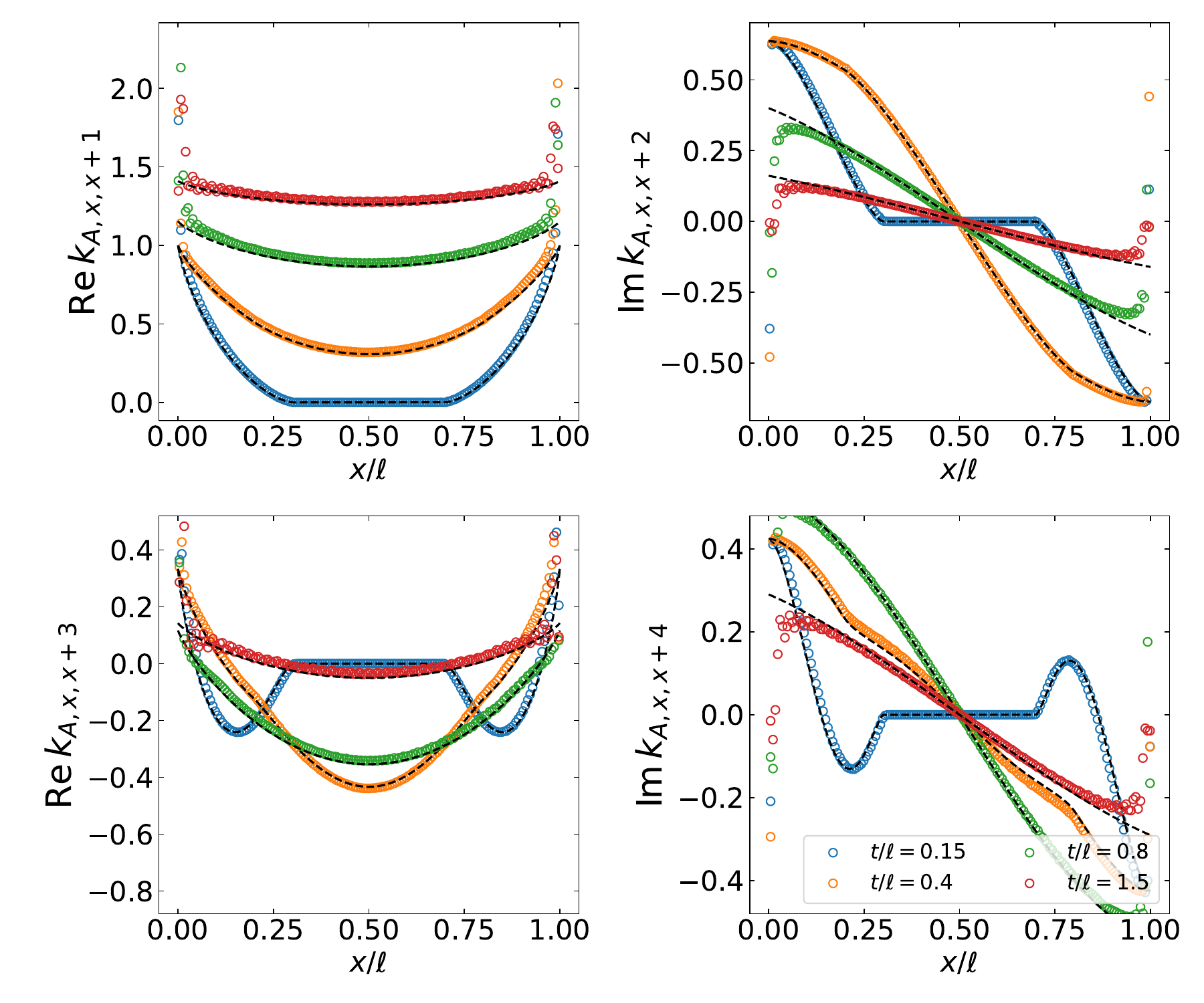}
    \caption{Quench from the staggered dimer state with $L_x=800$ and $L_y=8$, with separation $z_y=0$ and values of $z_x$ ranging from 1 to 4 in the four plots. The black dashed lines represent the quasiparticle prediction \eqref{eq:finalEH}}
    \label{fig:stdim}
\end{figure}

The second state considered is the staggered dimer, which is a simple modification of the collinear dimer. It is defined as
\begin{equation}
    \ket{\text{SD}} = \prod_{x+y=\text{even}} \frac{1}{\sqrt{2}}\left(c^\dagger_{x,y} -c^\dagger_{x+1,y}\right) \ket{0}.
\end{equation}
This initial state is more interesting as the correlation matrix has a nontrivial dependence on the degrees of freedom on the $y$-direction, and also a two-site shift symmetry along $y$. The correlation matrix elements are
\begin{equation}C^{SD}_{\boldsymbol{x},\boldsymbol{x}'} (t) = C^{\infty}-\frac{1}{L_y}\sum_{q_y} e^{iq_y(y-y')}(-1)^{y'}e^{-2it\cos q_y} \frac{i(x-x')}{4t}  e^{-i(x+x')\pi/2}J_{x-x'}(2t)
\end{equation} 
Where the asymptotic value $C^{\infty}$ is exactly the same as the one for the collinear dimer, showing that the long time results will be the same (recall that the Bessel function decays as $t^{-1/2}$ as $t \to \infty$).
However, the occupation functions are the same as in the collinear dimer,
\begin{equation}
    n_{SD}(\vk)= \frac{1}{2}(1-\cos(k_x))\rightarrow \eta_{SD}(k_x) = 2 \log \cot\frac{k_x}{2}.
\end{equation}
Therefore the quasiparticle prediction \eqref{eq:finalEH} will be the same. Hence the results imply that the $q_y$ dependence will have to be washed away in the strip geometry; indeed this is the case, as shown in figure \ref{fig:stdim}. In particular, the results are identical to the ones of the collinear dimer, as can be seen although they are plotted for different values of time.

\subsection{Diagonal Dimer}
The final comparison we consider is with the diagonal dimer state, defined as 
\begin{equation}
    \ket{\text{DD}} = \prod_{x=1}^{L_x/2}\prod_{y=1}^{L_y} \frac{1}{\sqrt{2}}\left(c^\dagger_{2x,y} -c^\dagger_{2x+1,y+1}\right) \ket{0}.
\end{equation}
Also in this situation the state exhibits a nontrivial structure along $y$, which enters through  coupling between sites at different $y$ positions,
\beq
 C_{\vx, \vx'}^{DD}(t) = C^{(\infty)} + \frac{\delta_{y+1,y'}}{4}i^{1-x-x'}J_{x-x'+1}(2t) +\frac{\delta_{y-1,y'}}{4}i^{1-x-x'}J_{x-x'-1}(2t)
\eeq
Here, however, the asymptotic value is now different, \beq
C^{(\infty)}= \frac{1}{2}\delta_{\vx, \vx'} -\frac{1}{4} (\delta_{x',x+1}\delta_{y',y+1}+\delta_{x',x-1}\delta_{y',y-1})
\eeq
Compared to the previous sections, this structure also affects significantly the quasiparticle prediction, since the occupation functions have nontrivial $k_x$ and $k_y$ dependence:
\begin{equation}
    n(\vk) = \frac{1}{2}(1-\cos(k_x+k_y)) \rightarrow \eta(\vk)=2\log \cot\frac{k_x+k_y}{2}
\end{equation}
\begin{figure}
    \centering
    \includegraphics[width=0.8\linewidth]{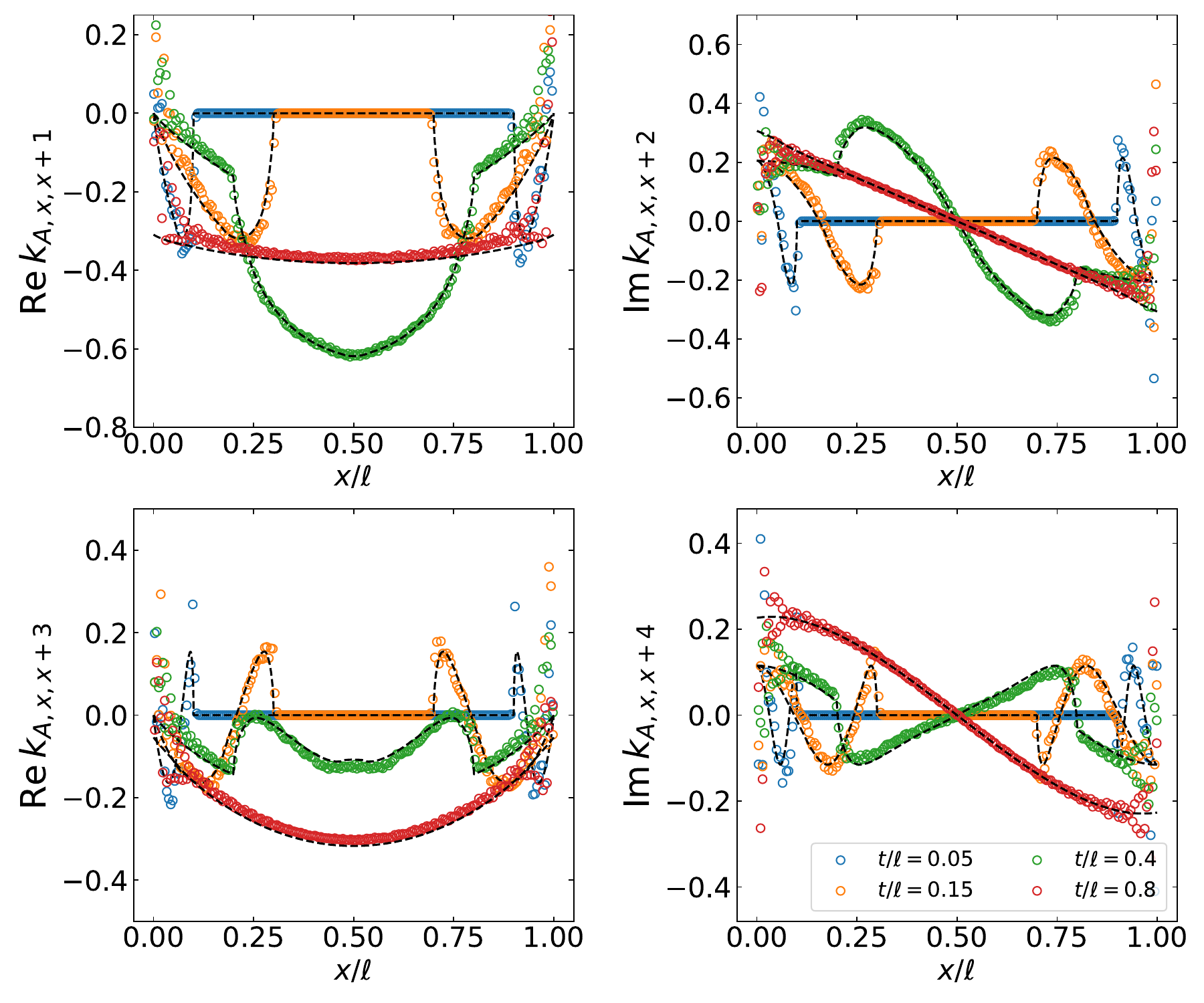}
    \caption{Quench from a diagonal diagonal dimer with $L_x=1000$ and $L_y=10$, with separation $z_y = 1$ and values of $z_x$ ranging from 1 to 4 in the four plots. }
    \label{fig:diagdimer}
\end{figure}
The comparison between the analytic and numerical results are shown in~\ref{fig:diagdimer} showing excellent agreement.

\subsection{A peculiar initial condition}
\label{sec:crossed}
As previewed earlier, in some special circumstances the standard QPP needs to be modified by performing a unitary transformation on the space of quasiparticles prior to evolving them in time. This was seen previously in the context of certain symmetry breaking states in one dimension~\cite{ares2023lack} but is not restricted to either one dimension or symmetry breaking states as we discuss in this section~\cite{DimensionalReduction}. The root cause in both cited examples  is the over abundance of conserved charges, which are present for free fermion systems on rectangular lattices, beyond the usual ones associated to the mode occupation, $\hat{n}(\boldsymbol{q})$. These extra conserved quantities then allow for a freedom in the choice of propagating quasiparticles which is fixed based upon properties of the initial state. 

In the two dimensional case, these extra quantities can be identified by first decomposing the Hamiltonian as
\begin{equation}
  H= H^X + H^Y,\hspace{0.5cm}  H^X = - \sum_{\vk} \cos k_x c_{\vk}^\dagger c_{\vk},\hspace{0.5cm}  H^Y = - \sum_{\vk} \cos k_y c_{\vk}^\dagger c_{\vk},
\end{equation}
where $H^X$ and $H^Y$ generate the motion in the $x$-direction and $y$-direction respectively and moreover $[H^X,H^Y]=0$.
Since the eigenvalues of $H^X$ do not depend on $k_y$, this Hamiltonian has a much larger class of integrals of motion compared to a generic 2D free Hamiltonian. In particular,
\begin{equation}
\left[H^X,c^\dagger_{k_x,k_y}c_{k_x,k_y'}\right]=0\hspace{0.3cm} \forall k_y,k_y'.
\end{equation}
Moreover, because of the geometry we consider, $H^X$ will determine which quasiparticles leave or enter the subsystem and therefore if they need to be traced out of the density matrix. Hence, if the initial state ``activates" one of these conserved charges, they will be constant along the motion which should be taken into account by the choice of quasiparticle. 

This is the case for the crossed dimer state,
\begin{equation}
 \ket{\text{C}} = \prod_{x=0}^{L_x/2-1}  \prod_{y=0}^{L_y/2-1} \frac{1}{2}(c_{2x,2y}^\dagger - c^\dagger_{2x+1,2y+1})(c_{2x+1,2y}^\dagger - c^\dagger_{2x,2y+1})\ket{0},
\end{equation}
which exhibits anomalous correlations of the form
\begin{equation}
\bra{C}c^\dagger_{k_x,k_y}c_{k_x,k_y+\pi}\ket{C} = -\frac{i}{2}\cos k_x \sin k_y.
\label{eq:extraconserved}
\end{equation}
As was shown in~\cite{DimensionalReduction} these lead to an unusual behaviour of the entanglement entropy which does not depend upon $n(\boldsymbol{k})$ in the expected manner. 

To determine the entanglement Hamiltonian for this state we proceed as before and divide the system into fluid cells of size $\boldsymbol{\Delta}=(\Delta,L_y)$,
\begin{align}
\ket{\psi}=\prod_{x_0=1}^{L/\Delta}\prod_{y=0}^{L_y/2-1}\prod_{p_x\in\frac{2\pi}{\Delta} \mathbb{Z}_\Delta}\sum_{k_x,k_x'=0,\pi}\frac{1}{16}(b_{x_0,q_x;2y}^\dagger - e^{-iq_x}b_{x_0,q_x;2y+1}^\dagger )(e^{-iq_x'}b_{x_0,q_x';2y}^\dagger - b_{x_0,q_x';2y+1}^\dagger )
\end{align}
where $q_x=p_x+k_x,~q_x=p_x+k_x'$ and  we have used the operators
\begin{eqnarray}
    b^\dag_{x_0,q_x;y}=\frac{1}{\sqrt{\Delta}}\sum_{\varkappa=0}^{\Delta-1}e^{-i\varkappa q_x}c^{\dag}_{x_0+\varkappa,y}
\end{eqnarray}
which are Fourier transformed along the $x$-direction only. We then rewrite this in terms of an appropriate basis of quasiparticles which respect the conservation of~\eqref{eq:extraconserved}. The correct choice is given by \cite{DimensionalReduction}
\begin{eqnarray}
   b_{x_0,q_x,y_+}^\dag =\frac{b^\dag_{x_0,q_x;2y}+ b^\dag_{x_0,q_x;2y+1}}{\sqrt{2}} \\
        b_{x_0,q_x;y_-}^\dag=\frac{b^\dag_{x_0,q_x;2y}-b^\dag_{x_0,q_x;2y+1}}{\sqrt{2}} 
\end{eqnarray}
which have mode operators $\hat{n}_{x_0,q_x;y_\pm}=b_{x_0,q_x,y_\pm}^\dag b_{x_0,q_x,y_\pm}$.  In this basis the initial state becomes
\begin{equation}
\ket{\psi}=\prod_{x_0=1}^{L/\Delta}\prod_{y=0}^{L_y/2-1}\prod_{p_x\in\frac{2\pi}{\Delta} \mathbb{Z}_\Delta}\sum_{k_x,k_x'=0,\pi}\Big[\sum_{\sigma=\pm}f_\sigma(q_x)b_{x_0,q_x;y_\sigma}^\dagger \Big]\Big[\sum_{\sigma=\pm}f_\sigma(q_x')b_{x_0,q'_x;y_\sigma}^\dagger \Big]
\end{equation}
where $f_+(q_x)=(1-e^{-iq_x})/4=f_{-}(q_x-\pi)$. This state can then be written as a density matrix, factorized over fluid cell  momenta $p_x$ and $y$ positions,
\begin{eqnarray} \rho(t)=\prod_{x_0=1}^{L/\Delta}\prod_{y=0}^{L_y/2-1}\prod_{p_x\in\frac{2\pi}{\Delta} \mathbb{Z}_\Delta} \rho_{x_0,p_x;y}(t).
\end{eqnarray}
The QPP evolution along the $x$-direction can now be implemented on the new quasiparticle operators,
\begin{eqnarray}
   e^{-iH^Xt} b^\dag_{x_0,q_x,y\pm}e^{iH^Xt}\approx b^\dag_{x_t(q_x),q_x,y\pm},
\end{eqnarray}
with $x_t(q_x)=x_0+v_x(q_x)t$. At this point, we can proceed  as before to evaluate the reduced density matrix after a time $t$. The density matrix splits into mixed and pure parts which receive contributions depending on whether all, some or no quasiparticles from a multiplet are present. For example, upon  tracing out the $k_x=\pi$ quasiparticles, we find, after some algebra,
\begin{eqnarray} \rho_{A,x_0,p_x;y}=\prod_{\sigma=\pm} n_{\sigma}(p_x)\hat{n}_{x_0,p_x,y_\sigma}+(1-n_{\sigma}(p_x))(1-\hat{n}_{x_0,p_x,y_\sigma})
\end{eqnarray}
with $ n_{\sigma}(p_x)=\bra{\psi}\hat{n}_{x_0,p_x,y_\sigma}\ket{\psi}=\frac{1}{2}(1\pm \cos (p_x))$. A similar expression is found after tracing out the $k_x=0$ particles instead.  Hence, we have exactly the same pair structure seen in the normal cases, with the difference that the occupation functions only have a dependence on the $x$-component of momentum. 

At this point, we have yet to apply the evolution along the $y$-direction. However because the subsystem consists of the entire $y$-direction this can be done after the trace over $\bar{A}$, giving, 
\begin{eqnarray}\nonumber
    \rho_{\rm mixed}(t)\!\!&=&\!\!e^{-iH_A^Yt}\prod_{p_x}\prod_{\{x_0|x_t(k_x)\in A \,\&\, x_t(k_x+\pi)\notin A \}}\hspace{-30pt}\rho_{A,x_0,p_x;y}(t)\hspace{-20pt}\prod_{\{x_0|x_t(k_x)\notin A \,\&\, x_t(k_x+\pi)\in A \}}\hspace{-35pt}\rho_{A,x_0,p_x+\pi;y}(t)\,e^{iH_A^Yt}\,,\\
    \rho_{\rm pure}(t)\!\!&=&\!\!e^{-iH_A^Yt}\prod_{p_x}\prod_{\{x_0|x_t(k_x)\in A \,\&\, x_t(k_x+\pi)\in A \}}\hspace{-20pt}\rho_{x_0,p_x;y}(t)\,e^{iH_A^Yt}\,,
\end{eqnarray}
where $H_A^Y$is the restriction of $H^Y$ to the subsystem $A$.  The Entanglement Hamiltonian is now just a slight modification of the previous formulae, with $K_{A,QP}{(t)}=e^{-iH^Y_At}\tilde{K}_{A,QP}{(t)}e^{iH^Y_At}$ and
\begin{eqnarray}\nonumber
    \tilde{K}_{A,QP}{(t)} &=& \sum_{y=1}^{L_y/2-1}\sum_{\sigma=\pm}\Bigg[\int_{q_x>0} \frac{dq_x}{2\pi} \eta_\sigma(q_x)\int_0^{\min(2v_x(q_x) t,\ell)} \hspace{-20pt}\hat{n}_{x,q_x;y_\sigma} 
\\&&\hspace{60pt}+\int_{q_x<0} \frac{dq_x}{2\pi} \eta_\sigma(q_x)\int_{\max(\ell-2|v_x(q_x)| t,0)}^\ell\hspace{-20pt}\hat{n}_{x,q_x;y_{\sigma}} \Bigg].
\end{eqnarray}
Here, $\eta_{\sigma}(q)=\log\frac{1-n_{\sigma}(q)}{n_{\sigma}(q)}$ is a function only of $q_x$.  While it still remains to perform the evolution along $y$ it is important to note that this is just a unitary transformation on the reduced density matrix. As a result this does not affect many quantities of interest, such as the entanglement entropy or full counting statistics of the charge.

Using the same method as before we are immediately led to following expression for the entropy,
\begin{equation}
    S_A(t)= \sum_{y=1}^{L_y/2}\sum_{\sigma=\pm} \int \frac{dq}{2\pi}\min(2|v_x(q)|t,\ell)[n_{\sigma}(q)\log n_{\sigma}(q)+(1-n_{\sigma}(q))\log (1-n_{\sigma}(q))]
\end{equation}
Which is identical to the result which was found in \cite{DimensionalReduction} thereby validating our expression for $\tilde{K}_{A,QP}(t)$. Similar expression for the R\'enyi entropy and full counting statistics can also be derived.  In appendix \ref{sec:AppA} we will discuss more general states in which this peculiar effect can arise, and we will explain its origin directly at the level of the correlation matrix in appendix \ref{sec:AppB}. 

\section{General geometries}
\label{sec:gengem}
In this section, we explore more complex scenarios where the choice of subsystem geometry does not allow for the application of dimensional reduction techniques.  
We shall once again focus on two-dimensional systems, but now take $L_x,L_y\to \infty$ and choose a subsystem which is extended in both $x$ and $y$ directions. The appropriate limit to consider in this case consists of taking $L_x,L_y\gg \sqrt{|A|}$, where $|A|$ is the volume of $A$, along with $L_x,L_y, \sqrt{|A|},t\to \infty$ and holding $t/\sqrt{|A|}$ fixed. This dictates our choice of fluid cell which we take to be rectangular with sides given by $\boldsymbol{\Delta}=(\Delta_x,\Delta_y)$. Accordingly, we split the lattice coordinate into $\boldsymbol{x}=\boldsymbol{x}_0+\boldsymbol{\varkappa}$ where $\boldsymbol{x}_0\in \mathbb{Z}^2_{\boldsymbol{L/\Delta}}$ denote the cell positions and $\boldsymbol{\varkappa}\in\mathbb{Z}^2_{\boldsymbol{\Delta}}$ the position inside the cell.

For simplicity we  return to using the symmetry breaking, squeezed states. Expressed in terms of the fluid cell modes they are
\beq
    \ket{\psi} = \mathcal{N}\exp{\Big(\sum_{\vx_0}\sum_{\vk}\mathcal{M}(\vk)b_{\vx_0,\vk}^\dagger b_{\vx_0,-\vk}^\dagger\Big)}\ket{0}
\eeq
where the fluid cell creation operators are 
\beq
 b_{\vx,\vk}^\dagger = \frac{1}{\sqrt{\Delta_x\Delta_y}}\sum_{\boldsymbol{\varkappa}}e^{-i\boldsymbol{\varkappa}\cdot\boldsymbol{k}}c^\dagger_{\vx_0 + \boldsymbol{\varkappa}}.
\eeq
The semiclassical evolution of these will still be of the form 
\begin{equation}
    e^{iHt} b^\dag_{\vx, \vk} e^{-iHt} = b^\dag_{\bold{x}_0 + \boldsymbol{v}(\vk)t,\vk}
\end{equation}
where now the both components of the velocity determine the motion. The analysis proceeds analogously to the one dimensional case. From the point of view of the quasiparticle picture, the only complication in this case is the counting of particles which are shared between $A$ and its complement. However, the basic idea is the same. In the following we show explicitly what would happen in the case of a simple geometry and then generalize to the generic case.

\subsection{Circle geometry}
Consider a subsystem which, in the coarse grained hydrodynamic limit, is a  circle of radius $R$ centered in the origin. 
Let us denote the positions of the quasiparticles by 
$
    \boldsymbol{x}_{\pm} = \boldsymbol{x}_0 \pm \boldsymbol{v}_{\vk}t
$
for $|\vk|>0$ with the $\vk$ dependence left implicit. 
The shared quasiparticles, where one resides within the circle and the other lies outside, are characterized by the set 
\begin{equation}
   \big\{\boldsymbol{x}_0\big | \,| \boldsymbol{x}_+|\leq R\,\&\, |\boldsymbol{x}_+ - 2\boldsymbol{v}_{\vk}t|>R, \forall \vk \big\}
\end{equation}
where we have expressed everything in terms of the $+$ coordinate, but accounted for all $\vk$, ensuring no overcounting. In this case, it is straightforward to see that the entanglement Hamiltonian becomes
\begin{equation}
    K_{A,QP}{(t)} = \sum_{\vk}\sum_{ \{\boldsymbol{x}_0| \,| \boldsymbol{x}_+|\leq R\,\&\, |\boldsymbol{x}_+ - 2\boldsymbol{v}_{\vk}t|>R\}}\eta(\vk)b^\dagger_{\vx_0,\vk}b_{\vx_0,\vk},
\end{equation}
This can then be transformed into real space by performing the inverse Fourier transform on the quasiparticles. Upon passing to the continuum description we arrive at 
\beq
K_{A,QP}{(t)}=\int d^2\vx \int_{\vx-\bold{z}\in A} \hspace{-15pt}d^2\bold{z}\,\mathcal{K}(\vx,\boldsymbol{z}) c^\dagger_{\vx}c_{\vx-\bold{z}}
\eeq
where the Kernel is given by 
\begin{equation}
    \mathcal{K}(\vx,\boldsymbol{z}) = \int \frac{d^2\vk} {(2\pi)^2} \eta(\vk) \Theta(|\vx|<R<|\vx-2\boldsymbol{v}_{\vk}t|)e^{-i\vk\cdot \boldsymbol{z}},
    \end{equation}
and we have used the notation 
\begin{equation}
     \Theta(|\vx|<R) = \begin{cases}
         1  \text{  if  } |\vx|<R\\
         0 \text{  otherwise  }
     \end{cases}.
\end{equation}
This has a natural and very intuitive structure compared to the one dimensional case: the argument of the kernel has to be a counting function which selects the shared quasiparticle pairs.

From this expression it is a simple matter to extract the evolution of the entanglement entropy. Using 
\beqa\nonumber
     \int d^2\vx \Theta(|\vx|<R)\Theta(|\vx-2\boldsymbol{v}_{\vk}t|>R) \!\!\! &=& \!\!\!\pi R^2 - \!\int d^2\vx \Theta(|\vx|<R)\Theta(|\vx-2\boldsymbol{v}_{\vk}t|<R) \\
     &=&\!\!\!2R^2 \arcsin\left(\frac{t|\boldsymbol{v}|}{R}\right) -2\frac{t|\boldsymbol{v}|}{R} \sqrt{1-\left(\frac{t|\boldsymbol{v}|}{R}\right)^2},
\eeqa
we find that the entanglement behaves as 
\begin{equation}
  S_A(t) = \int \frac{d^2 \vk}{2\pi^2}\Big[\!R^2 \arcsin\left(\frac{t|\boldsymbol{v}|}{R}\right) -\frac{t|\boldsymbol{v}|}{R} \sqrt{1-\left(\frac{t|\boldsymbol{v}|}{R}\right)^2}\Big]\![n(\boldsymbol{k})\log n(\boldsymbol{k})+(1-n(\boldsymbol{k}))\log (1-n(\boldsymbol{k}))]. 
\end{equation}
This can be checked to exhibit both an early time linear increase as well as late time saturation to a volume law term $\lim_{t\to\infty} S_A(t)\propto \,\pi R^2$ as expected.

 \subsection{General case}
 The generalization to arbitrary geometries is straightforward. Given an arbitrary connected region $A\subset \mathbb{\mathbb{Z}}^2$, we will have always the form
\begin{equation}
K_{A,QP}{(t)}=\int d^2\vx \int_{\vx-\bold{z}\in A} \hspace{-15pt}d^2\bold{z}\,\mathcal{K}(\vx,\boldsymbol{z}) c^\dagger_{\vx}c_{\vx-\bold{z}}
\end{equation}
Where the kernel is,
\begin{equation}
    \mathcal{K}(\vx,\boldsymbol{z})  = \int \frac{d^2\vk}{(2\pi)^2}\eta(\vk)e^{-i\vk\cdot \boldsymbol{z}}B(A,\vx,\boldsymbol{v}({\vk}),t)
    \label{fin}
\end{equation}
in which $B$ is the locus of points which satisfy $\vx \in A$ and $(\vx-2t\boldsymbol{v}(\vk))\notin A$, see, for example, the circular geometry of the previous subsection.
Note that this result does not  immediately appear to be consistent with the 1D one of \cite{1D}, expressed in terms of right and left movers. However, they coincide when Eq. \eqref{fin} is specialised to 1D. In that case, it is not difficult to see that 
\begin{equation}
B(\ell,x,v(k),t)= \Theta(\min(2v(k)t,\ell)-x)\delta_{v(k)>0} +  \Theta(\max(2v(k)t+\ell,0)+x)\delta_{v(k)<0}
\end{equation}
Which corresponds exactly to the result of \cite{1D}. 

\section{Conclusions}
\label{sec:concl}
In this paper we have obtained an effective description of the time-dependent entanglement Hamiltonian after a quantum quench in free fermionic systems in spatial dimensions  $d\geq 1$. 
For geometries in which dimensional reduction is applicable, we have shown that it is possible to obtain very general results that extend to a broad class of initial states: 
as long as there is an emerging pair structure in the $x$ direction, the results of \cite{1D} can be straightforwardly generalized to these settings.  
In more complicated geometries, the pair structure is a less common feature, and therefore the result for the squeezed state might seem rather a special case than the rule. 
However, it is straightforward to see that for more complex multiplets of quasiparticles in the initial state, the result does not break down completely. 
Instead, it requires a simple conceptual generalization of the above approach, involving a more intricate counting procedure to accommodate the additional complexity.

An intriguing generalization of the results presented here involves calculating the entanglement Hamiltonian when region  $A$  is not compact, such as when it consists of two disjoint subsets. While this extension would primarily require careful bookkeeping of quasiparticles without introducing significant conceptual challenges, it also represents an important step toward determining the negativity Hamiltonian \cite{murciano2022negativity,rottoli2023finite}. The negativity Hamiltonian provides an operator-level characterization of entanglement in mixed states, offering deeper insights into their structure and properties and so far it has been computed only in equilibrium settings.

The second natural line of research which will be pursued in the near future relates to the extensions of the result to interacting systems, in particular to integrable theories in 1+1 dimensions. Since these models feature stable quasiparticle excitations, it is natural to expect that some of the ideas of this work can be somehow extended also in such situations. However, the failure of the quasiparticle picture for the R\'enyi entropies in such models \cite{PRX} poses serious problems in this regard, and it is not clear whether some progress can be performed in this direction.

\medskip

 \noindent {\bf Acknowledgments:} 
We thank Filiberto Ares, Angelo Russotto and Federico Rottoli for useful discussions on the topic.
PC and CR acknowledge support from ERC under Consolidator Grant number 771536 (NEMO) and from European Union - NextGenerationEU, in the framework of the PRIN Project HIGHEST number 2022SJCKAH$\_$002.

\appendix
\section{General initial states}

\label{sec:AppA}
In this appendix, we investigate the quasiparticle structure for different classes of states of the form \eqref{eq:symm_preserv_general}. The simplest case to consider is given by
\begin{equation}
\ket{\Psi^{\boldsymbol{\mu}}}= \prod_{\boldsymbol{j}=1}^{\boldsymbol{L}/\boldsymbol{\mu}}(\sum_{\boldsymbol{m}=0}^{\boldsymbol{\mu}-1}a_{\boldsymbol{m}} c_{\boldsymbol{\mu} \boldsymbol{j}-\boldsymbol{m}}^\dagger)\ket{0}\hspace{1cm} \sum |a_{\boldsymbol{m}}|^2=1
\label{eq:oneparticle}
\end{equation}
 where we have used the compact notations 
 \begin{equation}
\prod_{\boldsymbol{j}=1}^{\boldsymbol{L}/\boldsymbol{\mu}} = \prod_{j_x=1}^{L_x/\mu_x}\prod_{j_y=1}^{L_y/\mu_y}\dots\prod_{j_d=1}^{L_d/\mu_d},\hspace{0.5cm} \sum_{\boldsymbol{m}=1}^{\boldsymbol{\mu}-1} = \sum_{m_x=1}^{\mu_x-1}\sum_{m_y=1}^{\mu_y-1} \dots
 \end{equation}
 and also we have expressed $\boldsymbol{\mu}\boldsymbol{j}=(\mu_xj_x, \mu_yj_y,...)$. We can expand this state over the basis of eigenstates of momentum to see the quasiparticle structure which arises. To do so we follow closely the derivation of \cite{molly}
 \begin{equation}
c^\dagger_{\boldsymbol{\mu}\boldsymbol{j}-\boldsymbol{m}}=\sum_{\boldsymbol{p}\in \frac{2\pi}{\boldsymbol{L}}\mathbb{Z}_{\boldsymbol{L}/\boldsymbol{\mu}}} e^{i\boldsymbol{p}\cdot (\boldsymbol{\mu}\boldsymbol{j}-\boldsymbol{m} )}\frac{1}{|\boldsymbol{\mu}|^{1/2}}\sum_{\boldsymbol{k}\in \frac{2\pi}{\boldsymbol\mu}\mathbb{Z}_{\boldsymbol{\mu}} }  e^{-i\boldsymbol{k} \boldsymbol{m}}c^\dagger_{\boldsymbol{p}+\boldsymbol{k}},
 \end{equation}
 inserting back into the state this gives
 \beqa
\ket{\Psi^{\boldsymbol{\mu}}}&=& \prod_{\boldsymbol{j}=1}^{\boldsymbol{L}/\boldsymbol{\mu}}\left[\sum_{\boldsymbol{m}=1}^{\boldsymbol{\mu}-1}a_{\boldsymbol{m}} \left(\sum_{\boldsymbol{p}\in \frac{2\pi}{\boldsymbol{L}}\mathbb{Z}_{\boldsymbol{L}/\boldsymbol{\mu}}} e^{i\boldsymbol{p}\cdot (\boldsymbol{\mu}\boldsymbol{j}-\boldsymbol{m} )}\frac{1}{|\boldsymbol{\mu}|^{1/2}}\sum_{\boldsymbol{k}\in \frac{2\pi}{\boldsymbol\mu}\mathbb{Z}_{\boldsymbol{\mu}} }  e^{-i\boldsymbol{k} \boldsymbol{m}}c^\dagger_{\boldsymbol{p}+\boldsymbol{k}}  \right)\right]\ket{0} \\
&=&  \prod_{\boldsymbol{j}=1}^{\boldsymbol{L}/\boldsymbol{\mu}} \sum_{\boldsymbol{p}} \sum_{\boldsymbol{m} } a_m e^{i\boldsymbol{p}\cdot (\boldsymbol{\mu}\boldsymbol{j}-\boldsymbol{m} )}\frac{1}{|\boldsymbol{\mu}|^{1/2}}\sum_{\boldsymbol{k}}  e^{-i\boldsymbol{k} \boldsymbol{m}}c^\dagger_{\boldsymbol{p}+\boldsymbol{k}} \ket{0}\\
&=& \prod_d \left[\det(e^{i\boldsymbol{p}_\alpha \boldsymbol{\mu}\boldsymbol{\beta}})_{\alpha,\beta\in \mathbb{Z}_{\boldsymbol{L}_d/\boldsymbol{\mu}_d}}\right] \prod_{\boldsymbol{p}\in \frac{2\pi}{\boldsymbol{L}}\mathbb{Z}_{\boldsymbol{L}/\boldsymbol{\mu}}} \frac{1}{|\boldsymbol{\mu}|^{1/2}}\sum_{\boldsymbol{m} } \sum_{\boldsymbol{k}}a_{\boldsymbol{m}} e^{-i(\boldsymbol{p} + \vk)\cdot \boldsymbol{m} }c^\dagger_{\boldsymbol{p}+\boldsymbol{k}} \ket{0}.\quad
 \eeqa
Note that, given $\tilde{k}=\prod_d\det(...)$, normalization of the state implies
\begin{equation}
    \braket{\Psi^\mu|\Psi^\mu}= \tilde{k}^2 \prod_{\boldsymbol{p}} \sum_{\boldsymbol{m}}|a_{\boldsymbol{m}}|^2 =1
\end{equation}
which sets $\tilde{k}=1$ by normalization. Hence we can write the state as
\begin{equation}
    \ket{\Psi^{\boldsymbol{\mu}}}= \prod_{ \boldsymbol{p}\in \frac{2\pi}{\boldsymbol{L}}\mathbb{Z}_{\boldsymbol{L}/\boldsymbol{\mu}}}  \ket{\psi_{\boldsymbol{p}}^{\boldsymbol{\mu}}} = \prod_{ \boldsymbol{p}\in \frac{2\pi}{\boldsymbol{L}}\mathbb{Z}_{\boldsymbol{L}/\boldsymbol{\mu}}}  \frac{1}{|\boldsymbol{\mu}|^{1/2}}\sum_{\boldsymbol{m} } \sum_{\boldsymbol{k}}a_{\boldsymbol{m}} e^{-i(\boldsymbol{p} + \vk)\cdot \boldsymbol{m} }c^\dagger_{\boldsymbol{p}+\boldsymbol{k}} \ket{0}.
\end{equation}
Introducing the object $f_{\boldsymbol{p}+\vk}=  \frac{1}{|\mu|^{1/2}}\sum_{\boldsymbol{m}}a_{\boldsymbol{m}}e^{i(\boldsymbol{p}+\vk)\boldsymbol{m}}$, the state describing the multiplet of correlated particles becomes
\begin{equation}
    \ket{\psi_{\boldsymbol{p}}^{\boldsymbol{\mu}}} = \sum_{\vk\in \frac{2\pi}{\boldsymbol{\mu}}\mathbb{Z}_{\boldsymbol{\mu}}} f_{\boldsymbol{p},\vk} c^\dagger_{\boldsymbol{p}+\vk}\ket{0},
    \label{eq:state}
\end{equation}
and the associated density matrix simply becomes
\begin{equation}
    \rho_p = \sum_{\vk}f_{\boldsymbol{p}+\vk} c^\dagger_{\boldsymbol{p}+\vk}\ket{0} \bra{0} \sum_{\vk'}f^*_{\boldsymbol{p}+\vk'} c_{\boldsymbol{p}+\vk'}
\end{equation}
where the vacuum is the one of the Hilbert space of the full sector which is obtained by varying $\vk$, namely of the Hilbert space $\mathcal{H}=\bigotimes_{\vk}\mathcal{H}_{\vk}$. Now, we note that the number of momenta in each direction is precisely equal to $\mu$. Since we consider dimensional reduction in which x is the only free axis, we require that $\mu_x=2$ to have quasiparticle structure along that direction, which we will refer to as + and - since their momenta are $k=0 \text{ and } \pi$, and since the velocity of the hopping model is a sine function $v(p+\pi)=-v(p)$, hence the two represent quasiparticles pairs moving with opposite velocities. Since, ultimately, we aim to trace out one of the two momenta and retain the other, it is not difficult to see that
\begin{equation}
    \rho_p = \frac{1}{|\mu|} \sum_{\vk}|f_{\boldsymbol{p}+\vk} |^2 c^\dagger_{\boldsymbol{p}+\vk}\ket{0} \bra{0} c_{\boldsymbol{p}+\vk} + \dots
\end{equation}
where the dots correspond to terms which give zero when tracing out, as can be easily seen because they have unbalanced creation and annihilation operators. As in the main text, we express the vacuum in terms of the number operators as
\begin{equation}
    \ket{0}\bra{0} = \prod_{\vk\in \frac{2\pi}{\boldsymbol{\mu}}\mathbb{Z}_{\boldsymbol{\mu}}} (1-\hat{n}_{\boldsymbol{p}+\vk}),
\end{equation}
and this leads to
\begin{equation}
    c^\dagger_{\boldsymbol{p}+\vk}\ket{0} \bra{0} c_{\boldsymbol{p}+\vk} = \hat{n}_{\boldsymbol{p}+\vk} \prod_{\boldsymbol{k}'\neq \vk} (1-\hat{n}_{\boldsymbol{p}+\vk'}),
\end{equation}
which means that 
\begin{equation}
    \rho_p =  \sum_{\vk}|f_{\boldsymbol{p},\vk} |^2 \hat{n}_{\boldsymbol{p}+\vk} \prod_{\boldsymbol{k}'\neq \vk} (1-\hat{n}_{\boldsymbol{p}+\vk'})  + \dots .
\end{equation}
At this point we can introduce the hydrodynamic approximation in strict analogy to the discussion of section \ref{sec:EhDimred}, by simply substituting $\hat{n}_{\vk + \boldsymbol{p}} \to\hat{n}_{\boldsymbol{x}_t(\boldsymbol{p}+\vk),\vk + \boldsymbol{p}} $, and a corresponding insertion of a tensor product over $x$. However for simplicity of notation we leave the
$x$ coordinate implicit, since it is obvious at this point that the right movers will have a position $x_0+v_x t$ and the left movers $x_0-v_xt$.
We also write $f^\pm_{\vk}$ and $\hat{n}^\pm_{\vk}$ to explicitly distinguish between the ones corresponding to the right mover and the left mover in the $x$ direction, assuming that $\vk$ refers to all momenta except for the one on the x direction. Assuming that $x_0+v_x t \in A$ while  $x_0-v_x t \in \overline{A}$, tracing out eliminates all the dependence on $n^-_{\vk}$ from the density matrix, and we obtain for the reduced density matrix of the right movers,
\beqa\nonumber
\rho_{\boldsymbol{p},A,+} &=& \sum_{\vk}|f^+_{\boldsymbol{\boldsymbol{p}+\vk}} |^2 \hat{n}^+_{\boldsymbol{p}+\vk} \prod_{\boldsymbol{k}'\neq \vk} (1-\hat{n}^+_{\boldsymbol{p}+\vk'})+ \frac{1}{|\boldsymbol{\mu}|} \sum_{\vk}|f^-_{\boldsymbol{\boldsymbol{p}+\vk}} |^2  \prod_{\boldsymbol{k}'} (1-\hat{n}^+_{\boldsymbol{p}+\vk'}) \\
   &=& \sum_{\vk} 
  \left(\prod_{\boldsymbol{k}'\neq \vk} (1-\hat{n}^+_{\boldsymbol{p}+\vk'})\right)\left(|f^+_{\boldsymbol{\vk}} |^2 \hat{n}^+_{\boldsymbol{p}+\vk} +|f_{\vk}^-|^2(1-\hat{n}^+_{\boldsymbol{p}+\vk})\right).
  \label{eq:weirdresult}
\eeqa
This expression is interesting in several aspects. First of all, we see that in 1D this reduces to the form already encountered several times in the main text,
\begin{equation}
    \rho_{p,+}^{1D}= |f^+|^2 \hat{n}^+ +|f^-|^2(1-\hat{n}^+),
\end{equation}
showing that in 1D the claim for the entanglement hamiltonian of the form \eqref{eq:finalEH} holds quite generally if $\mu_x=2$. Another case appears to lead immediately to the correct result, namely the one in which we have 1-site shift symmetry along all directions except for $x$. In this case there is only one accessible value of $k_y, k_z, \dots $ (which is 0), and therefore again
\begin{equation}
     \rho_{p,+}^{\mu_y, \mu_z, \dots =1}= |f^+_{0}|^2 \hat{n}^+ +|f^-_0|^2(1-\hat{n}^+).
\end{equation}
In other situations, however, the product in brackets becomes relevant. Its meaning is clear: the initial state we started from has a single excitation for each unit cell. Therefore, in the final GGE only one of the modes corresponding to the cell can be activated, and this product precisely project to zero all modes in the cell but one. Note that in section \ref{sec:EhDimred} this problem was not present because we considered states with 2-site shift symmetry along y and 2 excitations in each unit cell. It is not difficult to convince oneself that a result of the form \eqref{eq:itworks2} can be obtained for states with $|\boldsymbol{\mu}|/\mu_x=|\boldsymbol{\mu}|/2$ particles per unit cell, i.e. 

\begin{equation}
\ket{\Psi^{\boldsymbol{\mu}}}= \prod_{\boldsymbol{j}=1}^{\boldsymbol{L}/\boldsymbol{\mu}} \prod_{\lambda=1}^{|\boldsymbol{\mu}|/2}(\sum_{\boldsymbol{m}_\lambda=0}^{\boldsymbol{\mu}-1}a^{(\lambda)}_{\boldsymbol{m}_\lambda} c_{\boldsymbol{\mu} \boldsymbol{j}-\boldsymbol{m}_\lambda}^\dagger)\ket{0}.\hspace{1cm} 
\label{eq:versionwhichworks}
\end{equation}
For these states, repeating the argument above we always obtain  for the multiplet structure
\begin{equation}
     \rho_{\boldsymbol{p},A,+} = \prod_{\vk} 
\left(|F^+_{\boldsymbol{\boldsymbol{p}+\vk}} |^2 \hat{n}^+_{\boldsymbol{p}+\vk} +|F_{\boldsymbol{p}+\vk}^-|^2(1-\hat{n}^+_{\boldsymbol{p}+\vk})\right),
\label{eq:higher}
\end{equation}
where the $F_{\vk}$ are functions of the $f^{(\lambda)}_{\vk}$, which arise from the Kroenecker delta terms appearing in the same way as in section \ref{sec:EhDimred}, and can be related to the occupation functions of the state. 
As an example, we can go back to the two-dimensional states \eqref{eq:rhopfull}: integrating out the left moving component on the x direction, and introducing the notation $\hat{n}_{k_x,k_y}:=\hat{n}_{\vx_t(\boldsymbol{p}+\vk),\boldsymbol{p}+\vk}$ with $k_x,k_y= \pm$ to distinguish the two possible values, we obtain 
\beqa\nonumber
 \hspace{0.3cm}   \rho_{p,+} = \hat{n}_{++}\hat{n}_{+-}\left[F_{++}^{+-}+F^{++}_{+-}\right]+\hat{n}_{++}(1-\hat{n}_{+-})\left[F_{++}^{--}+ F^{++}_{--}+F^{-+}_{++}+F^{++}_{-+}\right]\hspace{1.3cm}
    \\ \nonumber  (1-\hat{n}_{++})\hat{n}_{+-}\left[F_{+-}^{--}+F^{+-}_{--}+F^{+-}_{-+}+F^{-+}_{+-}\right]+(1-\hat{n}_{++})(1-\hat{n}_{+-})\left[F_{--}^{-+}+F_{-+}^{--}\right].
\eeqa
It is easy to see that the $F_{\vk}$ functions are related to the expectation values,
\begin{equation}
    \left[F_{++}^{+-}+F^{++}_{+-}\right]=\braket{\hat{n}_{++}\hat{n}_{+-}},
\end{equation}
and since we consider gaussian initial states, we can apply Wick's theorem:
\begin{equation}
    \braket{\hat{n}_{++}\hat{n}_{+-}} = \braket{c^\dagger_{++}c_{++} }\braket{c^\dagger_{+-}c_{+-} }-\braket{c^\dagger_{++}c_{+-} }\braket{c^\dagger_{+-}c_{++} }\label{eq:terms}.
\end{equation}
To proceed we now assume that expectation values $\braket{c^\dagger_{++}c_{+-} }$ are zero: as will be explained below, this is related to the relaxation of the reduced density matrix in the long time limit. Therefore, all systems with this feature satisfy
\begin{equation}
     \braket{\hat{n}_{++}\hat{n}_{+-}} = \braket{\hat{n}_{++}} \braket{\hat{n}_{+-}} = n_{++} n_{+-},
     \label{eq:wick}
\end{equation}
and substituting this term in the above one immediately gets
\begin{equation}
   \rho_{\boldsymbol{p},A,+} = \prod_{k_y} \left( n_{+,k_y} \hat{n}_{+,k_y}+  (1-n_{+,k_y} )(1-\hat{n}_{+,k_y})\right).
   \label{eq:finalpairstructure}
\end{equation}
 Reintroducing the $\boldsymbol{p}$ dependence, this equation is rewritten as
\begin{equation}
      \rho_{\boldsymbol{p},A,+} = \prod_{k_y} \left( n_{p_x,p_y+k_y} \hat{n}_{p_x,p_y+k_y}+  (1-n_{p_x,p_y+k_y} )(1-\hat{n}_{p_x,p_y+k_y})\right)
   \label{eq:finalpairstructure}
\end{equation}
These kind of density matrices for the multiplets are precisely what is needed to obtain a result of the same form as the one of section \ref{sec:EhDimred}. In fact, it is easy to convince oneself that this tensor structure over $k_y$ combined with the product over $p_y$ and leads precisely to an entanglement hamitlonian \eqref{eq:finalEH}. The same can be done from all states of the form \eqref{eq:higher}, in any dimensionality.

One might now wander what structural difference is present between the two type of states considered, \eqref{eq:versionwhichworks} and \eqref{eq:oneparticle}, and why our discussion should apply just to the second type of states. 
Considering states of the form \eqref{eq:oneparticle}, it is easy to see that they satisfy $\braket{\hat{n}_{k_x,k_y}\hat{n}_{k_x,k_y'}}=0$. Using Wick theorem and considering only $d=2$ for simplicity this implies:
\begin{equation}
    \braket{c_{p_x,p_y+k_y}^\dagger c_{p_x,p_y+k_y'}}\neq 0\hspace{0.3cm} \forall k_y,k_y' \in \frac{2\pi}{\mu_y}\mathbb{Z}_{\mu_y}.
    \label{eq:norelaxationcondition}
\end{equation}
In \cite{DimensionalReduction}, it was proven that the reduced density matrix $\rho_A$ in a strip geometry relaxes to a stationary value $\rho_\infty$ in the long time limit if and only if all terms as \eqref{eq:norelaxationcondition} are zero. Therefore, the problem of a certain class of states is the fact that they do not thermalize: this is indeed what is implied by equation \eqref{eq:weirdresult}. Therefore the problem of states as \eqref{eq:oneparticle} is that they do not stationarize in the long time limit, and hence our discussion cannot be applied, since the quasiparticle picture cannot account for non-thermalizing dynamics: in this sense, they are analogs to the crossed dimer state considered in section \ref{sec:crossed}, and can be dealt with in the same way.  Note that this is a feature of the specific geometry considered, since the non-thermalizing part is always given by modes propagating in the y direction which remain trapped inside the strip. We will discuss precisely where the problem arises at the level of the correlation matrix in the end of the following appendix. Note finally that this feature is not exclusive of states \eqref{eq:weirdresult}, but could also happen in states as \eqref{eq:versionwhichworks} in special situations (as the crossed dimer).

\section{Exact results for some initial states}
\label{sec:AppB}
In this appendix we prove some of the exact expressions for the correlation matrix of the specific initial states considered in the text. These are simple two-dimensional generalizations of a class of calculations which is well known in 1D, see for instance \cite{Eisler_2007,parez2021exact}. For all the states considered, we use the conventions of \cite{DimensionalReduction}.
\subsection*{Collinear Dimer}
The correlation matrix of the initial state is given by:
\beqa\nonumber    \bra{CD}c_{\vx}^\dagger c_{\vx'}\ket{CD} &=& \frac{1}{2}\delta_{\vx,\vx'}-\frac{1}{4}\delta_{y,y'}\delta_{x+1,x'}[1-(-1)^{x}] - \frac{1}{4}\delta_{y,y'}\delta_{x-1,x'}[1+(-1)^{x}]\hspace{0.5cm} \\
    &=& \delta_{y,y'}\left(\frac{1}{2}\delta_{x,x'} -\frac{1}{4}\delta_{x+1,x'}[1-(-1)^{x}] - \frac{1}{4}\delta_{x-1,x'}[1+(-1)^{x}] \right).
\eeqa
Note that the term in brackets in the second line is exactly the correlation matrix for the 1D dimer state, which we will write $C^{1D}_{x,x'}$.
This immediately leads to the evaluation of $n(\vk)$
\begin{equation}
    n(\vk)= \bra{CD}c^\dagger_{\vk}c_{\vk}\ket{CD} = \frac{1}{2}(1-\cos(k_x)) \rightarrow \eta(k) = 2 \log \cot\left|\frac{k_x}{2} \right|,\label{eq:occcolldim}
\end{equation}
which is completely independent on the y component on momentum, which then will not be present at all in the entanglement Hamiltonian. From \eqref{eq:realspacecoeff_L} and \eqref{eq:realspacecoeff_R}, it is easy to see that the absence of any dependence on the y component of momentum gives rise to a factor $\delta_{y,y'}$. Hence the expected result from quasiparticle picture is
\begin{equation}
     K_{A,QP}(t) = \sum_{y} \int_0^l dx \int dz[ \mathcal{K}_R(x,z;t)+ \mathcal{K}_L 
 (x,z;t)]c_{x,y}^\dagger c_{x-z,y}.
\end{equation}
Note that the occupation functions are exactly the same as the ones of the 1D dimer which was studied in \cite{1D}.
Hence we expect that also the exact solution will be equivalent to the 1D solution if $y=y'$, and will be zero if this is not the case. This is not too difficult to see. Consider the correlation matrix in the mixed momentum-real space representation, namely 
\begin{equation}
    (C_{q_y})_{x,x'}(t)=\bra{\psi(t)}c^\dagger_{x,q_y}c_{x',q_y} \ket{\psi(t)},
\end{equation}
which is the partial Fourier transform along y of the usual correlation matrix $C_{\boldsymbol{x},\boldsymbol{x}'}$.
The specific form of the correlation matrix at time 0, and in particular the factorization of the term $\delta_{y,y'}$, allows to write:
\begin{equation}
     (C_{q_y})_{x,x'}(t) =C^{1D}_{x,x'}(t),
\end{equation}
where $C^{1D}_{x,x'}(t)$ is the correlation matrix of the 1D dimer state, which was studied in \cite{Eisler_2007}. It can be expressed in a compact form in terms of Bessel functions if we assume thermodynamic limit on the x axis,
\begin{equation}
    C^{1D}_{x,x'}(t)  = C^{\infty}-e^{-i\pi/2 (x+x')}\frac{i(x-x')}{4t}J_{x-x'}(2t),
\end{equation}
where $C^{\infty}=\frac{1}{2}\left(\delta_{x,x'} - \frac{1}{2}(\delta_{x,x'+1}+\delta_{x,x'-1})\right)$ is the correlation matrix at $t=\infty$.
Note that there is no actual dependence on $q_y$, and therefore when passing to real space over y we simply obtain a delta function,
\begin{equation}
    C_{\boldsymbol{x},\boldsymbol{x}'}(t)=\delta_{y,y'}(C_{q_y})_{x,x'}(t) = \delta_{y,y'}C^{1D}_{x,x'}(t).
\end{equation}
Therefore, we see that the prediction from the quasiparticle picture is exactly confirmed by analytical results. Note that for even $z=i_x-i_x'$ the element of the correlation matrix is imaginary, while for odd z the element is real: this justifies this choice in the plots of section \ref{sec:examples}

\subsection*{Staggered dimer}
This initial state has a $t=0$ correlation matrix of the form:
\begin{multline}
\bra{SD}c_{\vx}^\dagger c_{\vx'}\ket{SD} = \frac{1}{2}\delta_{\vx,\vx'}-\frac{1}{4}\delta_{y,y'}\delta_{x+1,x'}[1-(-1)^{x+y}] - \frac{1}{4}\delta_{y,y'}\delta_{x-1,x'}[1+(-1)^{x+y}] \\
=\delta_{y,y'}\left(\frac{1}{2}\delta_{x,x'} -\frac{1}{4}\delta_{x+1,x'}[1-(-1)^{x+y}] - \frac{1}{4}\delta_{x-1,x'}[1+(-1)^{x+y}] \right)
\end{multline}
This leads to modes:
\begin{equation}
    n(\vk) = \frac{1}{2}(1-\cos k_x) =  n(\vk)_{CD},
\end{equation}
which are the same as in the collinear dimer. Hence also in this case the quasiparticle prediction is greatly simplified. The correlation matrix however maintains a nontrivial y dependence: in momentum space,
\begin{equation}
    \bra{\text{SD}} c_{\vk}^\dagger c_{\vk'}\ket{\text{SD}}= \delta_{\vk,\vk'}\frac{1}{2}(1-\cos(k_x+k_y)) - \frac{i}{2}\delta_{k_y+\pi,k_y'} \delta_{k_x+\pi,k_x'}\sin(k_x+k_y).\label{eq:momentumsd}
\end{equation}
At this point the time-dependent real space correlation matrix can be evaluated noting that 
\begin{equation}
(C_{\boldsymbol{x},\boldsymbol{x}'}) (t)  = \frac{1}{L_x L_y} \sum_{\vk,\vk'}e^{i \vk \cdot \vx}e^{-i\vk' \cdot \vx'}e^{it(\varepsilon_{\vk}-\varepsilon_{\vk'})}\bra{\phi_0} c_{\vk}^\dagger c_{\vk'}\ket{\phi_0}
\label{eq:realspacecorrgeneral}
\end{equation}
for a generic initial state $\ket{\phi_0}$. Substituting expression \eqref{eq:momentumsd} this gives
\begin{equation}
    (C^{SD}_{\boldsymbol{x},\boldsymbol{x}'}) (t) = C^{\infty}-\frac{1}{L_y}\sum_{q_y} e^{iq_y(y-y')}(-1)^{y'}e^{-2it\cos q_y} \frac{i(x-x')}{4t}  e^{-i(x+x')\pi/2}J_{x-x'}(2t)
    \label{eq:stdimcorr}
\end{equation} 
where the term $C^{\infty}$ is the same as in the collinear dimer, and we have used that in the thermodynamic limit the sum over the x component of momentum reduces to a representation of a Bessel function:
\begin{equation}
    \int_{0}^{2\pi} \frac{dk}{2\pi} \cos(n k) e^{-2it \cos(k)} = i^{-n}J_n(2t).
\end{equation}
 The result cannot be simplified as we don't assume any thermodynamic limit on y. If that were the case, the integral over $q_y$ would just give a second Bessel function in terms of $y$ and $y'$.

\subsection*{Diagonal dimer}
The initial state correlation matrix is
\begin{equation}
      \bra{\text{DD}}c_{\vx}^\dagger c_{\vx'}\ket{\text{DD}} = \frac{1}{2}\delta_{\vx,\vx'}-\frac{1}{4}\delta_{y+1,y'}\delta_{x+1,x'}[1+(-1)^{x}] - \frac{1}{4}\delta_{y-1,y'}\delta_{x-1,x'}[1-(-1)^{x}]
\end{equation}
from which we get a similar result for the occupation functions,
\begin{equation}
    n(\vk) = \frac{1}{2}(1-\cos(k_x+k_y)) \rightarrow \eta(\vk)=\log \left(\cot\frac{k_x+k_y}{2}\right)^2.
\end{equation}
In momentum space this is immediately evaluated as
\begin{equation}
    \bra{\text{DD}} c_{\vk}^\dagger c_{\vk'}\ket{\text{DD}}= \delta_{\vk,\vk'}\frac{1}{2}(1-\cos(k_x+k_y))- \frac{i}{2}\delta_{k_y,k_y} \delta_{k_x+\pi,k_x'}\sin(k_x+k_y)
    \label{eq:initialdimer}
\end{equation}
The real space correlation matrix is then
\begin{equation}
    (C_{\vx, \vx'}^{DD})(t)= \frac{1}{L_xL_y} \sum_{\vk \vk'} e^{i \vk\cdot \vx} e^{-i \vk' \cdot \vx'} e^{it(\varepsilon_{\vk} -\varepsilon_{\vk'})} \braket{c_{\vk}^\dagger c_{\vk'}}_{DD},
\end{equation}
where the first term of \eqref{eq:initialdimer} contributes the asymptotic value of the correlation matrix
\beq
C^{(\infty)}= \frac{1}{2}\delta_{\vx, \vx'} -\frac{1}{4} (\delta_{x',x+1}\delta_{y',y+1}+\delta_{x',x-1}\delta_{y',y-1}),
\eeq
while the second term has a representation in terms of Bessel functions precisely as in the above:
\beqa\nonumber
 C_{\vx, \vx'}^{DD}(t) - C^{(\infty)} &=& -\frac{i}{2L_xL_y} \sum_{\vk, \vk'} e^{i \vk\cdot \vx} e^{-i \vk' \cdot \vx'} e^{it(\varepsilon_{\vk} -\varepsilon_{\vk'})}  \delta_{k_y,k_y} \delta_{k_x+\pi,k_x'}\sin(k_x+k_y) \\\nonumber
 &=& -\frac{1}{4L_x}\sum_{k_x}e^{ik_x(x-x')}(-1)^{x'} e^{-2it\cos k_x}\left[\delta_{y',y+1}e^{iq_x} -\delta_{y',y-1}e^{-iq_x}\right]\\
  &\approx&\frac{\delta_{y+1,y'}}{4}i^{1-x-x'}J_{x-x'+1}(2t) +\frac{\delta_{y-1,y'}}{4}i^{1-x-x'}J_{x-x'-1}(2t)
  \label{eq:diagdimcoll}
\eeqa
\subsection*{Crossed dimer and relaxation issues }
The crossed dimer is identified by the correlation matrix:
\begin{eqnarray}
\braket{C|c^\dagger_{\vx}c_{\vx'}|C} = \frac{\delta_{\vx,\vx'}}{2}-\frac{1}{8}\delta_{x+1,x'}\delta_{y+1,y'}\left(1+(-1)^{x}+(-1)^y+(-1)^{x+y}\right) \\ \nonumber
-\frac{1}{8} \delta_{x-1,x'}\delta_{y-1,y'}\left(1-(-1)^x-(-1)^y+(-1)^{x+y}\right) \\ \nonumber
-\frac{1}{8} \delta_{x-1,x'}\delta_{y+1,y'}\left(1-(-1)^x+(-1)^y-(-1)^{x+y}\right) \\ \nonumber
-\frac{1}{8} \delta_{x+1,x'}\delta_{y-1,y'}\left(1+(-1)^x-(-1)^y-(-1)^{x+y}\right)
\end{eqnarray}
In \cite{DimensionalReduction}, it was shown that this state is somehow problematic, since it does not lead to a stationary distribution $\rho^\infty$ in the long time limit. In this section we show that this effect can be diagnosed directly at the level of the correlation matrix and is strictly related to the discussed problems of states \eqref{eq:weirdresult}.
Fourier transforming to momentum space gives:
\beqa
    \braket{C|c^\dagger_{\vk}c_{\vk'}|C} =\frac{1}{2}\delta_{\vk,\vk'}(1-\cos k_x \cos k_y) - \frac{i}{2}\delta_{k_y,k_y'}\delta_{k_x+\pi,k_x'}\sin k_x \cos k_y \\ \nonumber -  \frac{i}{2}\delta_{k_y+\pi,k_y'}\delta_{k_x,k_x'}\cos k_x \sin k_y - \frac{1}{2}\delta_{k_x+\pi,k_x'}\delta_{k_y+\pi,k_y'}\sin k_x \sin k_y
\eeqa
At this point, the logic is analogous to the above. The first term gives the time independent part $C^{\infty}$. The terms with $\delta_{k_x+\pi,k_x'}$ give rise to Bessel functions in $x$ and $x'$. These behave nicely since the Bessel functions decay as $t^{-1/2}$ for large time. Hence all of such terms disappear in the long time limit. The problems arise in the first term of the second line: using \eqref{eq:realspacecorrgeneral} to obtain the real space correlation matrix this gives: 
\begin{equation}
    \frac{i}{4}\left[\delta_{x+1,x'}+\delta_{x-1,x'}\right]\frac{1}{L_y} \sum_{k_y}e^{ik_y(y-y')}(-1)^{y'} e^{-2it \cos k_y}\sin q_y.
    \label{eq:thirdterm}
\end{equation}
Where the sum does not converge as $t\to \infty$. This is ultimately a consequence of the strip geometry of the system, and is valid irrespective if $L_y$ is finite or is sent to infinity. In the first case, the sum itself keeps oscillating in time. In the second case, we can use the Bessel function representation shown above:
\begin{equation}
   \lim_{L_y \to \infty} \frac{1}{L_y} \sum_{k_y}e^{ik_y(y-y')}(-1)^{y'} e^{-2it \cos k_y}\sin q_y \propto \frac{y-y'}{4t}J_{y-y'}(2t) 
   \label{eq:nothermaliz}
\end{equation}
In this case the problem comes from having to consider arbitrary $y, y'\in \R$. All the terms in which $y-y' > t^{3/2}$ will lead to undamped oscillations which break thermalization.

Note that this discussion can be extended to all states in which the condition \eqref{eq:norelaxationcondition} applies. In all such situations the correlation matrix will contain terms with trivial $x$ dependence (given by a set of delta functions as in \eqref{eq:thirdterm}) while along y there will be sums as \eqref{eq:nothermaliz} which present undamped oscillations. This explains the absence of relaxation, which in turn invalidates the discussion of the main text for such initial states. In particular we see that for all states \eqref{eq:weirdresult}, we need to perform a discussion similar to that of section \eqref{sec:crossed} in order to obtain the Renyi entropies. 

\section{Dimensional reduction in $d \geq 3$}
\label{sec:AppC}
Although the focus of the discussion in the main text has been on two dimensional systems, one can also straightforwardly consider higher dimensions. As discussed in appendix \ref{sec:AppA}, it is important to be careful about which states to consider, as in dimensional reduction geometries there might be several instances of initial states not leading to a GGE. We therefore consider states as \eqref{eq:versionwhichworks},
\begin{equation}
\ket{\Psi^{\boldsymbol{\mu}}}= \prod_{\boldsymbol{j}=1}^{\boldsymbol{L}/\boldsymbol{\mu}} \prod_{\lambda=0}^{|\boldsymbol{\mu}|/2}(\sum_{\boldsymbol{m}_\lambda=0}^{\boldsymbol{\mu}-1}a_{\boldsymbol{m}_\lambda} c_{\boldsymbol{\mu} \boldsymbol{j}-\boldsymbol{m}_\lambda}^\dagger)\ket{0}\hspace{1cm} .
\end{equation}
Given that such states in general lead to reduced density matrix of the form \eqref{eq:higher}, we simply obtain 
\beqa
 \rho^{(t)}_{A,R} &=& \prod_{p_x>0} \prod_{x=0}^{\min(2v_x(p_x)t,L_x)}  \prod_{\boldsymbol{p} \in  \frac{2\pi}{\boldsymbol{L}}\mathbb{Z}_{\boldsymbol{L}}} \rho_{p,x,+}^{(p_x,\boldsymbol{p})}, \\
 \rho^{(t)}_{A,L} &=&  \prod_{p_x<0} \prod_{x=\max(L_x-2|v_x(p_x)|t,0)}^{L_x} \prod_{\boldsymbol{p} \in  \frac{2\pi}{\boldsymbol{L}}\mathbb{Z}_{\boldsymbol{L}}} \rho_{p,x,+}^{(p_x,\boldsymbol{p})},
\eeqa
where $\boldsymbol{L}$ and $\boldsymbol{p}$ encompass all coordinates except for x. Hence we get immediately the (obvious) generalization of the entanglement hamiltonian
\beqa
     K_{A,QP}(t) &=&\sum_{\boldsymbol{i}} \sum_{\boldsymbol{\delta}} \int_0^l dx \int dz_x[ \mathcal{K}^{\boldsymbol{\delta}}_L(x,z_x;t)+ \mathcal{K}^{\boldsymbol{\delta}}_R(x,z_x;t)]c_{x,\boldsymbol{i}}^\dagger c_{x-z_x,\boldsymbol{i}- \boldsymbol{\delta}}
     \label{eq:finalEHmultidim},
\eeqa
where $i,\delta$ are (d-1)-dimensional vectors and
\begin{equation}
    \mathcal{K}^{\boldsymbol{\delta}}_{L,R} = \frac{1}{|\boldsymbol{L}|}\sum_{\boldsymbol{p}}e^{-i\boldsymbol{p}\cdot\boldsymbol{\delta}}\int_{q_x>0}\frac{dq_x}{2\pi}\eta(\boldsymbol{p}) \Theta_{L,R} e^{-iq_x z_x} 
\end{equation}
is the multidimensional generalization for the kernels, where we have expressed compactly $\Theta_L =  \Theta(\max(l+2v_x(q_x)t,0)+x)$ and $\Theta_R = \Theta(\min(2v_x(q_x)t,l)-x)$.

\bibliographystyle{ytphys}
\bibliography{bibliography}

\end{document}